\documentclass{article}
\title{Radiative Response of Atomic Systems Illuminated with Approximate Spherical Vector Waves}
\author{F. Camas-Aquino and R. J\'auregui}
\date{\today}

\usepackage{subcaption}

\usepackage{amsmath}
\usepackage{amssymb}
\usepackage{graphicx}
\usepackage{amsfonts}
\usepackage{euscript}

\usepackage{xcolor}

\begin{document}

\maketitle

\begin{abstract}
The natural electromagnetic modes spontaneously emitted by an atom in free space are spherical vector waves (SVWs). Each SVW mode is uniquely linked to a specific dynamical--spherical--multipole--moment of the atomic system. In this work, we introduce a general formalism for evaluating spherical multipole transition rates under different boundary conditions, considering the superposition between a given SVW and the modes resulting from the boundary conditions. This formalism is applied to study the radiative properties of an atomic system trapped near the focus of a 4$\pi$ optical array. By appropriately selecting the external light field, the juxtaposed lenses of the optical array allow the atom to be illuminated with approximate spherical vector waves. Explicit expressions for the resulting multipole transition rates are presented as a function of the numerical aperture of the lenses. The feasibility of enhancing and inhibiting electric dipole-forbidden transitions using such an array, under current experimental capabilities, is briefly discussed. 
\end{abstract}

\section{Introduction}

The implementation of reliable quantum engineering processes requires the coherent control of the interaction between tailored photons and individual atoms. 
Typical devices employ low-temperature atomic systems with preselected location and internal state \cite{Grimm2000,thompson2013}, and optical arrays that generate electromagnetic (EM) fields which, near the localized atomic system, exhibit intensity, phase and polarization profiles that result in strong light-matter coupling~\cite{endres16,Barredo2016}; this allows the manipulation of the global quantum state \cite{Kaufman2021,Langen2024}. Simultaneously, the properties of the EM field in the radiation region are expected to provide a precise characterization of the light-matter state.

We study the response of a single atom to light fields  whose structure resembles that of photons naturally radiated by  atoms in free space, i. e., spherical electromagnetic waves (SVW)~\cite{Bereteskii71}.  As a direct consequence of the relevance of the central field approximation to describe isolated atoms, and the validity of a spherical-multipole description of the transitions between internal atomic states, it is expected that 
SVW could improve the efficient control of photon-atom interactions, even beyond the standard electric dipole scheme. Our theoretical analysis incorporates the ideas behind  well established experimental devices used for the  generation of classical EM waves with a 3D subwavelength structure. The optical scheme, known as a 4$\pi$ array~\cite{Hell94}, is based on the  convergence of  external illumination at a common focus using juxtaposed lenses, Fig.~\ref{fig:4piOA_diagram}. The outcoming constructive interference is known to induce an axial resolution to a value below the  wavelength of the incoming fields.
Back in 2017, a 4$\pi$ array with incoming Gaussian beams was used to study the electromagnetic  response of a single atom located nearby the focus of the array \cite{Kurtsiefer2017}. It was found that the 4$\pi$  super-resolution imaging technique yielded a significant increase on the efficiency coupling of light to a single atom. Modified photon statistics of the transmitted field indicated a nonlinear interaction at the single-photon level.

Recently,  the feasibility of using a $4\pi$ array to create focused fields 
with controlled polarization and phase structures similar to SVW was reported~\cite{aquino2023}.
Here, we analize  how a single photon generated in the $4\pi$ array, suitable for creating approximate spherical vector waves, can be useful for inducing or inhibiting arbitrary multipole atomic transitions.

  The outlook of this manuscript is as follows. Section 2 gives a compact  description of electromagnetic wave sources in terms of SVWs. The one-to-one correspondence of multipole sources with specific SVWs is emphasized. The next section describes a formalism that incorporates boundary conditions in the electromagnetic field. These conditions make the usage of alternative basis sets necessary to describe the electrodynamics evolution of an atomic system. A specific and illustrative realization of this scenario is worked out in Section 4 in detail; there, approximate spherical waves are assumed to be generated by a 4$\pi$ array. These waves have an angular spectrum given by a clipped version of that of a SVW. The transition rates of atoms characterized by their spherical multipole moments and located near the focus of the 4$\pi$ array are evaluated. The results are discussed in the final section.

\section{Radiative Emission and Absorption from a Spherical Multipole Source}

The structure of atomic systems and their interaction with electromagnetic waves is determined by the elementary electrodynamic interaction operator,
\begin{equation}\label{eq:EMInteOper}
\hat{\mathcal{V}}(t) = \frac{1}{c} \int d^{3}{\mathbf{r}} \hat{J}^{\mu}(\mathbf{r},t)\hat{A}_{\mu}(\mathbf{r},t),\quad \mu =0,1,2,3.
\end{equation}
Here $\hat{J}^{\mu}(\mathbf{r},t)$ denotes the operator associated with the electric charge current density, and $\hat{A}=(c\hat{\Phi},\hat{\mathbf{A}})$  the electromagnetic field potential operator; Einstein summation notation is used.
In the Coulomb gauge $$\hat{\Phi}(\mathbf{r},t)=0,\quad \nabla\cdot\hat{\mathbf{A}}(\mathbf{r},t)=0,$$
and the electric $\hat{\mathbf{ E}}(\mathbf{r},t)$ and magnetic  $\hat{\mathbf{ B}}(\mathbf{r},t)$ fields are given by
$$ \hat{\mathbf{ E}}(\mathbf{r},t)=-\frac{1}{c}\frac{\partial\hat{\mathbf{ A}}(\mathbf{r},t)}{\partial t},\quad
\hat{\mathbf{ B}}(\mathbf{r},t) =\nabla\times \hat{\mathbf{ A}}(\mathbf{r},t).$$
The current density operator $\hat{J}^{\mu}(\mathbf{r},t)$ can be written in terms of a  multipole expansion with a symmetry selected to facilitate the interpretation of both the atomic system and the
EM field dynamics under given boundary conditions. An essential mathematical tool in a multipole expansion is the completeness relation satisfied by selected vector fields. Being now interested in atoms radiating into free space, we now focus on spherical vector waves to describe the field. 

\subsection{Spherical multipole expansion of the transition current $\mathbf{j}_{fi}(\mathbf{r})$}
For an atomic system, the electric charge current density is a bilinear operator of the matter field $\hat \Psi$,
\begin{equation}
\hat{J}_\mu = \hat \Psi^\dagger \hat{j}_\mu \hat \Psi.
\end{equation}
In the center of mass description of an electron with charge $e$ in a hydrogenic atom, the operator $\hat{j}_\mu$ corresponds to $e$ multiplied by the product of Dirac matrices $ \gamma_\mu$ with $\gamma_0$ in the relativistic regime, that is,
\begin{equation}
\hat{J}_\mu = e\hat \Psi^\dagger \gamma_0\gamma_\mu \hat \Psi;
\end{equation}
while, in the nonrelativistic regime, $\hat{j}_0$ is the charge $e$ times the identity operator,
\begin{equation}
\hat{J}_0 = e\hat \Psi^\dagger \hat \Psi
\end{equation}
and  $\hat{\mathbf{j}}$ is a differential operator,
\begin{equation}
\hat{\mathbf{J}} =  -\frac{ie\hbar}{2m}(\hat \Psi^\dagger \boldsymbol{\nabla}\hat \Psi - (\boldsymbol{\nabla}\hat \Psi^\dagger)\hat \Psi).
\end{equation}

In general, $\hat{J}_\mu (\mathbf{r},t)$ can be written as a linear combination of charge current density matrix with elements $j_{ba;\mu} (\mathbf{r})$ involving the stationary states of the atomic Hamiltonian with labels $a,b$,
\begin{eqnarray}
\hat{J}_\mu (\mathbf{r},t) &=&
 \sum_{a,b}  j_{ba;\mu} (\mathbf{r})  e^{-i(\omega_a-\omega_b)t} \hat b^\dagger_b \hat b_a,\\
 j_{ba;\mu} (\mathbf{r}) &=& \psi_b^\dagger(\mathbf{r}) \hat{j}_\mu\psi_a(\mathbf{r}).
\end{eqnarray}
Within first order perturbation theory, $j_{ba;\mu} (\mathbf{r})$  describes the energy shifts and transition probabilities induced by the electromagnetic field from an atomic stationary state $\vert a\rangle$ with energy $\hbar\omega_a$ to another atomic stationary state $\vert b\rangle$ with energy $\hbar\omega_b$.

Under standard conditions, the wavelength of radiative transitions is much larger than the size of the atomic system, and a multipole description of  $ j_{ba;\mu} $ is adequate. The success of the central field model in describing isolated atoms makes it natural to select spherical vector waves (SVWs) centered at the atomic center of mass to perform a symmetry adapted multipole expansion of any particular transition current element $j_{fi;\mu} (\mathbf{r})$ to analize the radiative properties of an 
atom~\cite{Bereteskii71}.

 In the Coulomb gauge, the vector potential of a monochromatic  SVW of angular frequency $\omega$, amplitude $\mathcal{A}_\nu$,
polarization $P$, total angular momentum $j$, and projection along the $z$ axis $m$ is 
\begin{equation}
\mathbf{A}_{\omega\nu}(\mathbf{r},t) = \mathcal{A}_\nu e^{-i\omega t} \mathbb{Y}_{k j m}^{(P)}(\mathbf{r}),
\quad\quad k=\frac{\omega}{c}.
\label{eq:Anu}
\end{equation}\\
To simplify the notation, the index $\nu$, introduced in the right hand side of this equation, encompasses the labels $P$, $j$, and $m$, that is $\nu=\{P,j,m\}$.  The polarization $P$ is called electric $E$, magnetic $M$, or longitudinal $L$ for reasons that will become clear later.
The vector spherical harmonics $\mathbb{Y}_{k j m}^{(P)}(\mathbf{r})$ have the angular spectrum representation,
\begin{equation} 
\label{eq:PlaneWaveDeco}
\mathbb{Y}_{k j m}^{(P)}(\mathbf{r}) = \frac{1}{(2\pi)^{3/2}}\int d\Omega_{\mathbf{n}} \tilde{\mathbb{Y}}_{jm}^{(P)}(\mathbf{n}) e^{ikr\mathbf{n}\cdot\mathbf{u}}, \quad \mathbf{n} = \frac{\mathbf{k}}{k}, \mathbf{u} =\frac{\mathbf{r}}{r}
\end{equation}
that depends on the polarization $P$ as follows,
\begin{subequations}\label{eq:SVW_AngSpe}
\begin{equation}\label{eq:ElecModes}
\tilde{\mathbb{Y}}_{jm}^{(E)}(\theta_\mathbf{n},\phi_\mathbf{n}) = \frac{1}{\sqrt{j(j+1)}} \nabla_{\mathbf{n}} Y_{jm} (\theta_\mathbf{n},\phi_\mathbf{n}),
\end{equation} 
\begin{equation}\label{eq:MagnModes}
\tilde{\mathbb{Y}}_{jm}^{(M)}(\theta_\mathbf{n},\phi_\mathbf{n}) = \mathbf{n}\times \tilde{\mathbb{Y}}_{jm}^{(E)}(\theta_\mathbf{n},\phi_\mathbf{n}),
\end{equation}
\begin{equation}\label{eq:LongModes}
\tilde{\mathbb{Y}}_{jm}^{(L)}(\theta_\mathbf{n},\phi_\mathbf{n}) = \mathbf{n} Y_{jm} (\theta_\mathbf{n},\phi_\mathbf{n}).
\end{equation}
\end{subequations}
$Y_{jm}(\theta_\mathbf{n},\phi_\mathbf{n})$ are the scalar spherical harmonics  
and $\nabla_{\mathbf{n}}$ is the angular part of the gradient operator in the wave vector domain $\nabla_{\mathbf{k}} = \mathbf{n}\partial_{k}+k^{-1}\nabla_{\mathbf{n}}$, 
\begin{equation}
\nabla_{\mathbf{n}} = \hat{\mathbf{e}}_{\theta_\mathbf{n}} \frac{\partial}{\partial\theta_\mathbf{n}} + \hat{\mathbf{e}}_{\phi_\mathbf{n}} \frac{1}{\sin{\theta_\mathbf{n}}} \frac{\partial}{\partial\phi_\mathbf{n}}.
\end{equation}
The spectrum representations of the electric  and magnetic fields of different polarizations $E$ and $M$ are directly related,
\begin{equation}
\tilde{\mathbf{E}}_{kjm}^{(E)} = -\tilde{\mathbf{B}}_{kjm}^{(M)}, \quad \tilde{\mathbf{B}}_{kjm}^{(E)}  =\tilde{\mathbf{E}}_{kjm}^{(M)}.\label{eq:EBBE}
\end{equation}

The explicit expressions for $\mathbb{Y}_{k j m}^{(P)}(\mathbf{r})$ in configuration space involve the spherical Bessel functions
$$g_\ell(kr) = \sqrt{\frac{\pi}{2kr}}J_{\ell +1/2}(kr)$$ as a consequence of the equation
\begin{equation}
\int d\Omega_\mathbf{n} e^{ikr\mathbf{n}\cdot\mathbf{u}} Y_{jm} (\mathbf{n}) = 4\pi i^{j} g_j(kr)  Y_{jm} (\mathbf{u}).
\end{equation}

The orthonormality of the angular spectrum functions $\tilde{\mathbb{Y}}_{jm}^{(P)}(\mathbf{n})$ can be directly verified,
\begin{equation}\label{eq:sportho}
\int d\Omega_{\mathbf{n}}\,\tilde{\mathbb{Y}}_{jm}^{(P)*}(\mathbf{n})\cdot\tilde{\mathbb{Y}}_{j^\prime m^\prime}^{(P^\prime)}(\mathbf{n}) = \delta_{jj^\prime}\delta_{mm^\prime}\delta_{PP^\prime}.
\end{equation}
They guarantee a similar behavior of the vector spherical harmonics 
\begin{equation}
\int_{\mathbb{R}^3} d^3{\mathbf{r}}\mathbb{Y}_{k^\prime j^\prime m^\prime}^{(P^\prime)*}(\mathbf{r})\cdot
\mathbb{Y}_{k j m}^{(P)}(\mathbf{r}) =\delta_{PP^\prime}\delta_{jj^\prime}\delta_{mm^\prime}\frac{\delta(k-k^\prime)}{k^2}.\label{eq:Yortho}
\end{equation}
Vector spherical harmonics also constitute a complete set, so that the components $p,q=$ $1,2,3$ of the identity tensor can be written as
\begin{equation}\label{eq:complete}
\Big[ \overleftrightarrow {\delta}(\mathbf{r} - \mathbf{r}^\prime) \Big]_{pq} =\int dk \,k^2\sum_{P,j,m} \Big[\mathbb{Y}_{k j m}^{(P)}(\mathbf{r})\Big]_p\Big[\mathbb{Y}_{k j m}^{(P)*}(\mathbf{r}^\prime)\Big]_q.
\end{equation}
As a consequence, the spatial components of the transition current ${\mathbf j}_{fi} (\mathbf{r})$ between given initial $i$ and final $f$ states can be written in terms of SVW,
\begin{eqnarray}
[\mathbf{j}_{fi}(\mathbf{r})]_p &=& \int d^3\mathbf {r}^\prime [\overleftrightarrow {\delta}(\mathbf{r} - \mathbf{r}^\prime)]_{pq}  [\mathbf{j}_{fi} ( \mathbf {r}^\prime )]_q\\
&=& \int dk\,k^{2}\sum_{\nu} [\mathbb{Y}_{k\nu}(\mathbf{r})]_p \big(Q_{\nu}(k)\big)_{fi}\label{eq:CurrDensSpheMultMomeDeco3}.
\end{eqnarray}
The  dynamic spherical multipole moment of order $\nu$ 
of the transition current density $\mathbf{j}_{fi}(\mathbf{r})$ is
\begin{equation}
\big(Q_{\nu}(k)\big)_{fi} = \int d^{3}\mathbf {r}^\prime\,\mathbf{j}_{fi}(\mathbf{r}^\prime)\cdot\mathbb{Y}_{k\nu}^{*}(\mathbf{r}^\prime).
\end{equation}

In the static limit $k\rightarrow 0$ the  dynamic spherical multipole moments reduce to the standard static  electric and magnetic multipoles for the $E$ and $M$ polarizations, respectively \cite{Jackson75}.
This becomes evident when several identities are applied. We begin with the $E$ polarization, where
$$
\mathbf{j}_{fi}(\mathbf{r})\cdot\mathbb{Y}_{kjm}^{(E)*}(\mathbf{r}) = i^{1-j}\sqrt{\frac{2}{\pi j(j+1)}}\Big[
\frac{1}{k}\Big[\nabla_{\mathbf{r}} \cdot\Big(\big(1 + r\frac{\partial}{\partial r}\big)g_j(kr)Y^*_{jm}(\mathbf{u}) \mathbf{j}_{fi}(\mathbf{r})\Big)$$
\begin{equation}
 -\Big( \Big( 1 + r\frac{\partial}{\partial r}\Big)g_j(kr)Y^*_{jm}(\mathbf{u})
 \nabla_{\mathbf{r}}\cdot \mathbf{j}_{fi}(\mathbf{r})\Big] + kr\mathbf{u}\cdot\mathbf{j}_{fi}(\mathbf{r}) g_j(kr)Y^*_{jm}(\mathbf{u})  \Big].
\end{equation}
Therefore, the continuity equation for a localized current leads to 
$$\big(Q^{(E)}_{jm}(k)\big)_{fi} =$$
\begin{equation}
i^{-j}\sqrt{\frac{2}{\pi j(j+1)}}\int d^3{\mathbf{r}}\Big[c\rho(\mathbf{r})
\Big(1 + r\frac{\partial}{\partial r}\Big)g_j(kr) + i \mathbf{u}\cdot\mathbf{j}_{fi}(\mathbf{r})krg_j(kr)\Big] Y^*_{jm}(\mathbf{u}).
\label{eq:SeM}
\end{equation}
The results follow for the $M$ polarization, where a similar analysis leads to
\begin{equation}
\big(Q^{(M)}_{jm}(k)\big)_{fi} =i^{-j}\sqrt{\frac{2}{\pi j(j+1)}}\int d^3{\mathbf{r}}\big(\mathbf{r}\times \mathbf{j}_{fi}(\mathbf{r})\big)\cdot\nabla_{\mathbf{r}}\Big(g_j(kr) Y^*_{jm}(\mathbf{u})\Big).\label{eq:SmM}
\end{equation}
Finally for the longitudinal polarization,
\begin{equation}
\big(Q^{(L)}_{jm}(k)\big)_{fi}=
i^{-j}c\sqrt{\frac{2}{\pi}} \int d^3{\mathbf r}\rho(\mathbf{r})g_j(kr)Y^*_{j m}(\mathbf{u}).
\end{equation} 
The well known static expressions for the spherical multipole moments result from recognizing that, for distances to the origin $r$ much smaller than the wavelength involved,
\begin{equation}
g_\ell(kr)\approx \frac{(kr)^\ell}{(2\ell +1)!!},\quad\quad kr\ll 1.
\end{equation}

The expressions Eq.~(\ref{eq:SeM}) and Eq.~(\ref{eq:SmM}) for the dynamic electric and magnetic multipole moments  respectively are used in high precision atomic calculations of transition probabilities \cite{Labzowsky93}-\cite{Bilal2019}. It is important to realize that relativistic expressions of the transition currents based on Dirac-like equations
already contain spin effects as clearly shown via the Gordon decomposition \cite{Greiner87}.

\subsection{Photons emitted and absorbed by spherical multipole sources}

We have shown that a transition current density of an atomic system $\mathbf{j}_{fi}$  can be naturally written using the spherical multipole expansion.
Likewise, the electromagnetic operator can then be expanded using the monochromatic spherical modes  $\mathbf{A}_{\omega\nu}$ defined in Eq.(\ref{eq:Anu}), resulting
\begin{equation}
\hat{\mathbf{A}}(\mathbf{r},t) =\sum_{\omega,\nu} \Big[\mathbf{A}_{\omega\nu}(\mathbf{r})e^{-i\omega t}\hat{a}_{\omega\nu}+\mathbf{A}^*_{\omega\nu}(\mathbf{r})e^{i\omega t}\hat{a}^\dagger_{\omega\nu}\Big].
\end{equation}
Here $\hat{a}_{\omega\nu}$ and $\hat{a}^\dagger_{\omega\nu}$ denote the standard annihilation and creation operators corresponding to the $\bar{\nu} =\{\omega \nu\}$-mode.
The SVW modes are orthonormalized according to Einstein rule,
$$\sum_{\bar{\nu}}  \hbar\omega (\hat{a}^\dagger_{\omega\nu} \hat{a}_{\omega\nu} +1/2)=
 $$
  \begin{equation}\frac{1}{8\pi} \int d^3{\mathbf{r}}\big( \hat{\mathbf{E}}( \mathbf{r},t) \cdot \hat{\mathbf{E}}( \mathbf{r},t)
+  \hat{\mathbf{B}}( \mathbf{r},t) \cdot \hat{\mathbf{B}}( \mathbf{r},t)\big).
\end{equation}
Explicitly, in the Coulomb gauge,
\begin{equation}\label{eq:EMVecPotComPmode1}
\mathbf{A}^{(P)}_{\omega j m}(\mathbf{r}) = \sqrt{\frac{hc^2}{\omega}} \mathbb{Y}^{(P)}_{k j m}(\mathbf{r}),\quad P=E,M.
\end{equation}
Longitudinal spherical vector modes are not involved in the expression of $\mathbf{A}_{\bar\nu}$ in such a gauge since the
divergence of those modes is not zero. 

Under standard conditions, the absorption and emission of single photons by a matter source can
be properly described using a perturbative treatment of the interaction operator Eq.~(\ref{eq:EMInteOper}). 
 The matrix element of the QED interaction operator between two states $|i\rangle$ and $|f\rangle$ along with the emission~(-) or absorption~(+) of a $\bar\nu$-photon is
\begin{equation}\label{eq:TMEt}
\mathcal{V}_{fi}(t)= e^{-i(\omega_{fi}\mp\omega)t} V_{fi},
\end{equation}
where $\hbar\omega_{fi}$ is the energy difference between $f$ and $i$ states and $V_{fi}$ is the time-independent transition matrix element.
Assuming that matter is illuminated by a specific SVW mode $\bar\nu_{0}=\{\omega_{0},P_{0},j_{0},m_{0}\}$, the corresponding time-independent transition matrix element is
\begin{equation}\label{eq:TME_SVW}
V_{fi}^{\bar\nu_{0}} =  -\sqrt{\frac{h}{\omega_0}} \Upsilon_{fi}^{\bar\nu_{0}} \Big(Q_{j_{0}m_{0}}^{(P_{0})}(k_{0})\Big)_{fi}.
\end{equation}
This expression is based on the fact that SVW's are orthonormal; the labels of the electromagnetic modes coincide with those of the  multipole moment of the transition current, reflecting the conservation of angular momentum and parity.
 The factor
$\Upsilon_{fi}^{\bar\nu_{0}}$ depends on the initial $I$ and final $F$ states of the EM field 
\begin{equation}
\Upsilon_{fi}^{\bar\nu_{0}} = \langle F\vert a^\dagger_{\bar \nu_0} \vert I\rangle
\end{equation}
for an emission process and
\begin{equation}
\Upsilon_{fi}^{\bar\nu_{0}} = (-1)^{-m_{0}}\langle F\vert a_{\bar \nu_0} \vert I\rangle
\end{equation}
for an absorption process. 

Illuminating matter with a specific SVW mode $\nu_{0}$ activates only one spherical multipole moment with the same labels as the illumination, i.e., $\nu^\prime=\nu_{0}$, which is the only mode able to perform the transition. Conversely, it is also true that a matter-specific spheric multipole moment can only generate the emission or absorption of a specific SVW mode with the same labels as the matter's multipole.

The time-independent factor associated to the transition probability for photon emission from a material source under illumination of a SVW $\bar\nu_{0}$-mode is
\begin{equation}
w_{fi}^{\bar\nu_{0}} =  \frac{4\pi^{2}}{\omega_{0}} \Big|\Upsilon_{fi}^{\omega_{0}\nu_{0}} \big(Q_{\nu_{0}}(k_{0})\big)_{fi}\Big|^{2}.
\end{equation}
The time dependent factor of the transition probability naturally depends on the relation between the illuminating source frequency $\omega_0$ to the frequency derived from the
atomic energy differences $\omega_f -\omega_i$ of atomic states $i$ and $f,$ and their natural line-widths.

\section{An Atomic System in an Engineered EM Environment}\label{sec:snu}

Up to now, we have studied the electrodynamic response of an atom in otherwise free space. However, the electromagnetic environment surrounding the atom can be  modified by means of diverse optical elements such as lenses, cavities and waveguides.
If the localization of such elements does not vary in the time scales of interest, the electromagnetic field operator is then naturally described using properly normalized monochromatic modes that incorporate the corresponding boundary conditions on the EM field, 
\begin{equation}
\hat{\mathbf{A}}(\mathbf{r},t) =\sum_{\bar{\sigma}} \Big[\boldsymbol{\mathcal{A}}_{\bar{\sigma}}(\mathbf{r})e^{-i\omega t}\hat{a}_{\bar{\sigma}}+\boldsymbol{\mathcal{A}}^*_{\bar{\sigma}}(\mathbf{r})e^{i\omega t}\hat{a}^\dagger_{\bar{\sigma}}\Big].
\end{equation}
Here $\bar\sigma=\{k,\sigma\}$ denotes the labels that identify each mode including its  wave vector modulus $k =\omega/c$.   
For an atomic system with spatial dimensions much smaller than the relevant wavelengths of radiative transitions, it is still convenient to preserve the spherical multipole description of the transition current, Eq.~(\ref{eq:CurrDensSpheMultMomeDeco3}).
In such a case, the complete character of SVW, Eq.~(\ref{eq:complete}) allows to write for each $q$ component of the vector $\boldsymbol{\mathcal{A}}_{\bar{\sigma}}(\mathbf{r})$ in terms of SVWs,
 \begin{eqnarray}
 \big[\boldsymbol{\mathcal{A}}_{\bar{\sigma}}(\mathbf{r})]_q &=& \int dk\,k^2\sum_\nu [\mathbb{Y}_{k\nu}(\mathbf{r})]_q \mathcal{R}(k;\sigma,\nu)\label{eq:sup}\\
  \mathcal{R}(k;\sigma,\nu) &=& \int d^3{\mathbf{r}}^\prime\boldsymbol{\mathcal{A}}_{\bar{\sigma}}(\mathbf{r}^\prime)\cdot\mathbb{Y}^*_{k\nu}(\mathbf{r}^\prime);
 \end{eqnarray} 
If the region of integration is described approximately by $\mathbb{R}^3$,
\begin{equation}
\mathcal{R}(k;\sigma,\nu) = \Big[\frac{\delta(k-\omega_{\bar{\sigma}}/c)}{k^2}\Big]\mathcal{O}(k;\sigma,\nu),
\end{equation}
 with the term $\mathcal{O}(k;\sigma,\nu)$ given directly by the angular spectrum $\widetilde{\boldsymbol{\mathcal{A}}}_{\bar{\sigma}}(\mathbf{n})$  of the $\bar{\sigma}$-modes,
\begin{equation}
\mathcal{O}(k;\sigma,\nu) =\int d\Omega_{\mathbf{n}}\tilde{\mathbb{Y}}^{(P)*}_{jm}(\mathbf{n})\cdot\widetilde{\boldsymbol{\mathcal{A}}}_{k\sigma}(\mathbf{n}).\label{eq:oksn}
\end{equation}
The $\bar{\sigma}$-modes are the eigenfunctions of the total --orbital plus polarization-- angular momentum
of the field. Then, $\mathcal{O}(k;\sigma,\nu)$ provides a quantitative measurement of the optical angular momentum of   an ${\mathcal{A}}_{\bar{\sigma}}$ mode of frequency $\omega = kc$.
Within a perturbative analysis, the relevant time-independent transition matrix element for the emission of photons associated to a mode $\boldsymbol{\mathcal{A}}_{\bar{\sigma}}$ is given by
\begin{equation}
V^{\bar{\sigma}}_{fi} = \int dk\, k^2\sum_\nu V^{\bar{\nu}}_{fi} \mathcal{R}(k;\sigma,\nu)
\end{equation}
and
\begin{equation}\label{eq:wfisn}
w_{fi}^{\bar{\sigma}_{0}} =  \frac{4\pi^{2}}{\omega_{0}}  \Big|\sum_{\nu}\Upsilon_{fi}^{\bar{\nu}}\mathcal{O}(k_0;\sigma_0,\nu) \big(Q_{\nu}(k_0)\big)_{fi}\Big|^{2}.
\end{equation}
$\mathcal{O}(k;\sigma_0,\nu)$ can be regarded as the analogue of the density of modes used in the evaluation of radiative transition rates in a description that uses EM plane-waves. The last equation shows that the boundary conditions that define the $\sigma$-modes may alter the transition rate for a given $\nu$-multipole of the transition current. 

\section{Approximate Spherical Vector Waves (ASVW) Generated by a $4\pi$ Optical Array}

The analysis presented in the previous section shows that even a single photon may  induce a given multipole transition with a probability that can be manipulated by selecting the EM vector wave properties, Eq.~\ref{eq:wfisn}. The location of the atomic system must be taken into account.
Previously  \cite{aquino2023},  the feasibility of using a $4\pi$ array to create focused fields 
with controlled polarization and phase structures similar to SVW was reported. In this Section we describe the expected atomic response to the presence of those modes if the atom is located close to the focus of the $4\pi$ array.

\begin{figure*}[hbt!]
    \centering
    \includegraphics[width=0.8\textwidth]{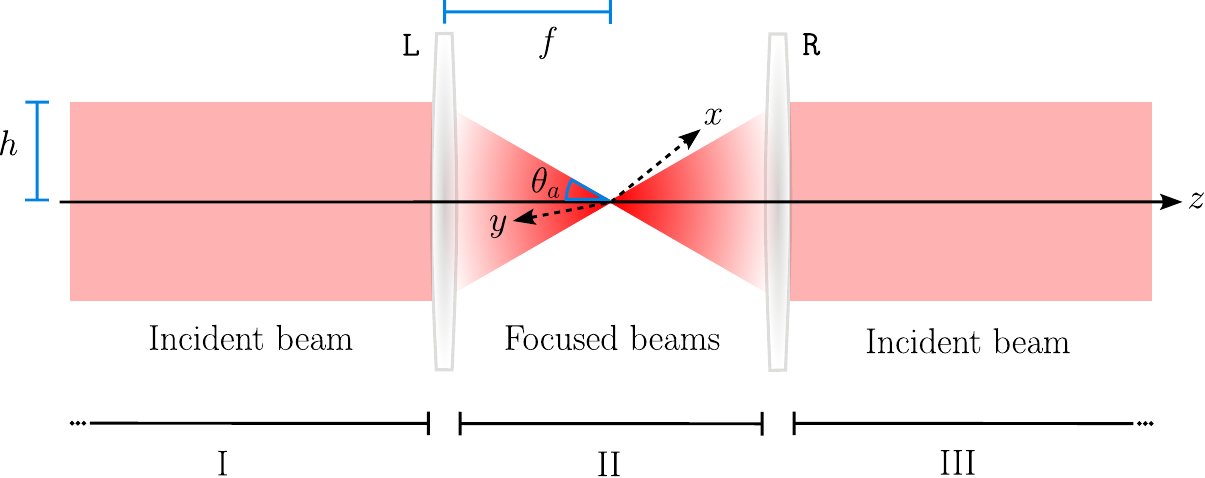}
    \caption{Schematics of a $4\pi$ opical array made of two aplanatic lenses $\mathtt{L}$ and $\mathtt{R}$: the external regions I and III cover the illumination of the lenses by collimated beams, and the internal region II covers their focusing.}
    \label{fig:4piOA_diagram}
\end{figure*}

 Standard confocal optical microscopy accomplishes
already very small probe volumes, nevertheless the excitation light that is focused
through a lens can cover at most half of the solid angle,
limiting the axial resolution due to a focal volume elongated along
the optical axis. This limitation has been overcome by using two
opposing lenses with coinciding focal point, known as 4$\pi$
array~\cite{Hell94}. The path of the incident beam is split, and the
object is coherently illuminated by two counter-propagating parts
of the ﬁeld simultaneously, Fig.~\ref{fig:4piOA_diagram}. In this way, the input mode
covers almost the entire solid angle, limited only by the numerical
aperture of the focusing lenses.

A given 4$\pi$ optical array defines three  spatial regions as illustrated in Fig.~\ref{fig:4piOA_diagram}.
The incoming EM fields in regions I and III  are  taken as cylindrical vector beams with
 their symmetry axis coinciding with that of the 4$\pi$ system. They are focused by the pair of
aplanatic lenses of usually high numerical aperture {\rm NA}, and their radial and azimuthal polarizations are chosen as described in Ref.~\cite{aquino2023}. Then,
in Region II, an approximate monochromatic spherical vector wave (ASVW)  
\begin{equation}
\mathbf{A}^{({\rm NA})}_{\bar{\nu}}(\mathbf{r},t)= \mathcal{A}_\nu e^{-i\omega t} \mathbb{Y}_{\omega j m}^{(P;{\rm NA})}(\mathbf{r}),
\quad\quad \nu=\{P,j,m\}
\label{eq:AnuNA}
\end{equation}
is generated with an angular spectrum representation
\begin{equation}
\label{eq:PlaneWaveDeco}
\mathbb{Y}_{k j m}^{(P;{\rm NA})}(\mathbf{r}) 
= \frac{1}{(2\pi)^{3/2}}\int_{\Omega_{{\rm NA}}}  d\Omega_{\mathbf{n}}
 \tilde{\mathbb{Y}}_{jm}^{(P)}(\mathbf{n}) e^{ikr\mathbf{n}\cdot\mathbf{u}}.
\end{equation}
The region of angular integration  $\Omega_{\rm NA}$ is delimited by the azimuthal angle $\phi\in [0,2\pi)$ and the polar angle $\theta\in [0,\theta_{{\rm NA}}]\cup [\pi-\theta_{{\rm NA}},\pi]\}$ with $\sin\theta_{\rm NA} =\mathrm{NA}$,
see Fig.~\ref{fig:4piOA_diagram}.
This expression for the field is valid only within the internal region II of the 4$\pi$ optical array.
Nevertheless, in the following we assume that the volume of region II, $\mathcal{V}_{II}$ (typically cm$^3$), is  large compared to the volume associated to the relevant wavelength, $\lambda^3 =(2\pi)^3/k^3$ (typically $\mu$m$^3$), so that
\begin{equation}
\frac{1}{(2\pi)^3} \int_{\mathcal{V}_{II}} d^3{\mathbf{r}} e^{i(\mathbf{k} -\mathbf{k}^\prime)\cdot\mathbf{r}} \approx \delta (\mathbf{k} -\mathbf{k}^\prime).
\end{equation}
Under these conditions, the overlap between a pair  $\mathbb{Y}_{k j m}^{(P;{\rm NA})}$ modes in configuration space can be evaluated in terms of their overlap in wave-vector space,
\begin{equation}\label{eq:CutedAngulInteg}
\mathcal{Y}_{\nu}^{\nu'}(\theta_{{\rm NA}})=\int_{\Omega_{\rm NA}} d\Omega_{\mathbf{n}}\,\tilde{\mathbb{Y}}_{\nu}(\mathbf{n})\cdot\tilde{\mathbb{Y}}_{\nu'}^{*}(\mathbf{n}).
\end{equation}
In general, the constraint in integration region for the polar angle $\theta_{\mathbf{n}}$ prevents the orthogonality of the $\mathbb{Y}_{\omega j m}^{(P;{\rm NA})}(\mathbf{r})$ modes. Calculating
$\mathcal{Y}_{\nu}^{\nu'}(\theta_{\rm NA})$
results in analytical expressions involving the 3-$j$ symbols, 
\begin{itemize}
\item $P=P'$, $P\in\{E,M\}$
\begin{multline}\label{eq:CutedAngulInteg_MM}
\mathcal{Y}_{Pjm}^{Pj'm'}(\theta_{\rm NA}) = \delta_{mm'} (-1)^{j+j'} 
\sqrt{(2j+1)(2j'+1)}
\\
\sum_{\lambda=\pm1,0} \mathcal{C}_{jj'}^{m+\lambda} (\theta_{\rm NA})
\begin{pmatrix}
j & 1 & j\\
m+\lambda & -\lambda & -m
\end{pmatrix}
\begin{pmatrix}
j' & 1 & j'\\
m+\lambda & -\lambda & -m
\end{pmatrix}
.
\end{multline}

\item $P\neq P'$
If $P=E$ and $P'=M$ or vice versa 
\begin{multline}
\mathcal{Y}_{Ejm}^{Mj'm'}(\theta_{\rm NA}) = \delta_{mm'} \sqrt{2j'+1} 
\sum_{\lambda=\pm1,0}
\Bigg\{
\sqrt{j}
 \mathcal{C}_{j+1,j'}^{m+\lambda} (\theta_{\rm NA})
\begin{pmatrix}
j+1 & 1 & j\\
m+\lambda & -\lambda & -m
\end{pmatrix}
\\
-\sqrt{j+1}\mathcal{C}_{j-1,j'}^{m+\lambda} (\theta_{\rm NA})
\begin{pmatrix}
j-1 & 1 & j\\
m+\lambda & -\lambda & -m
\end{pmatrix}
\Bigg\}
\begin{pmatrix}
j' & 1 & j'\\
m+\lambda & -\lambda & -m
\end{pmatrix}
\label{eq:YEM}.
\end{multline}
\item $P\in\{E,M\}$ and $P'=L$, $\mathcal{Y}_{Pjm}^{Lj'm'}(\theta_{\rm NA})
=0$ due to the orthogonality between the vectors in the integrand. 
\end{itemize}
In these equations, the dependence on the numerical aperture NA of the lenses is encoded in the coefficient 
$\mathcal{C}_{jj'}^{m+\lambda}(\theta_{\rm NA})$,
\begin{multline}\label{eq:OrthCoefASVWExp}
\mathcal{C}_{jj'}^{m+\lambda}(\theta_{\rm NA})=\frac{1+(-1)^{j+j^\prime}}{2}
i^{j-j^\prime}\sqrt{(2j+1)\frac{(j-|m+\lambda|)!}{(j+|m+\lambda|)!} (2j'+1)\frac{(j'-|m+\lambda|)!}{(j'+|m+\lambda|)!}}\\
\int\limits_{\cos(\theta_{\rm NA})}^{1}dxP_{j}^{|m+\lambda|}(x)P_{j'}^{|m+\lambda|}(x),
\end{multline}
This coefficient is zero when the parities of $j$ and $j'$ are different. The orthonormality property, Eq.~(\ref{eq:Yortho}), is recovered  when the numerical aperture is unit ($\theta_{\rm NA}=\pi/2)$.

\begin{figure*}[hbt!]
     \centering
\begin{tabular}{c c c}   
         \includegraphics[width=0.32\textwidth,trim={2cm 0.8cm 1.9cm 0.8cm},clip]{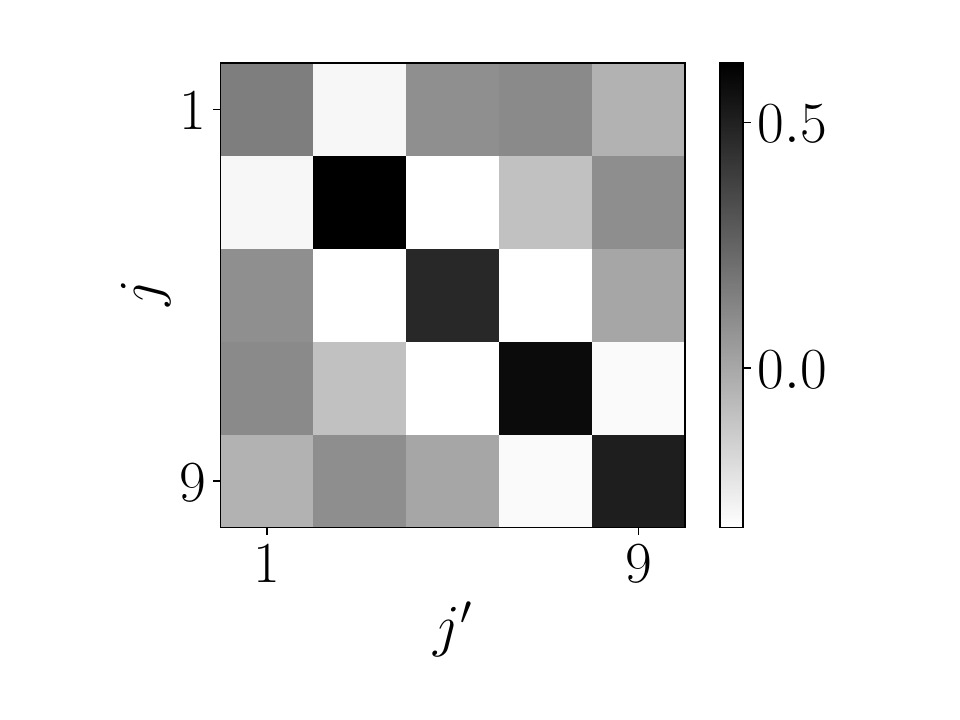}&
         \includegraphics[width=0.32\textwidth,trim={1.7cm 0.8cm 2.2cm 0.8cm},clip]{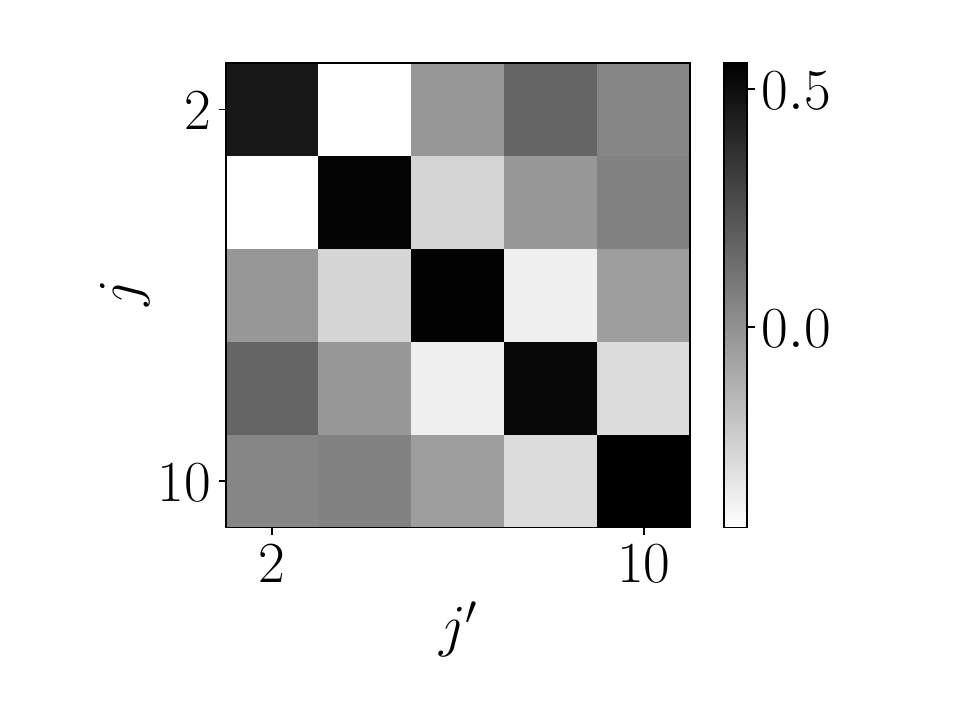}&
         \includegraphics[width=0.32\textwidth,trim={2.7cm 0.8cm 1.2cm 0.8cm},clip]{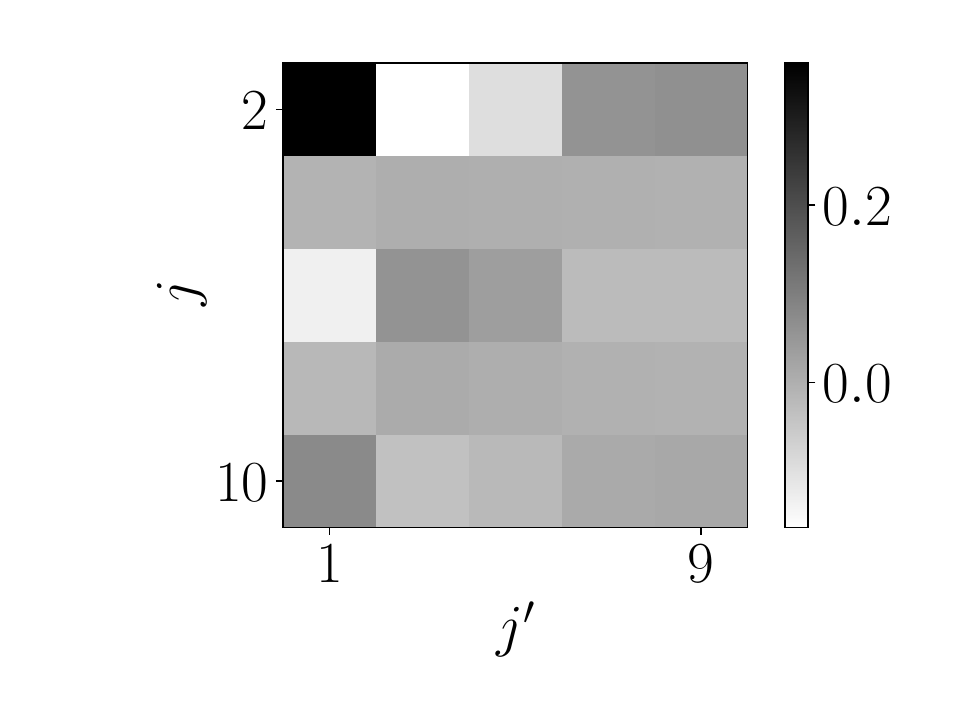}\\
         \includegraphics[width=0.32\textwidth,trim={3cm 0.8cm 0.9cm 0.8cm},clip]{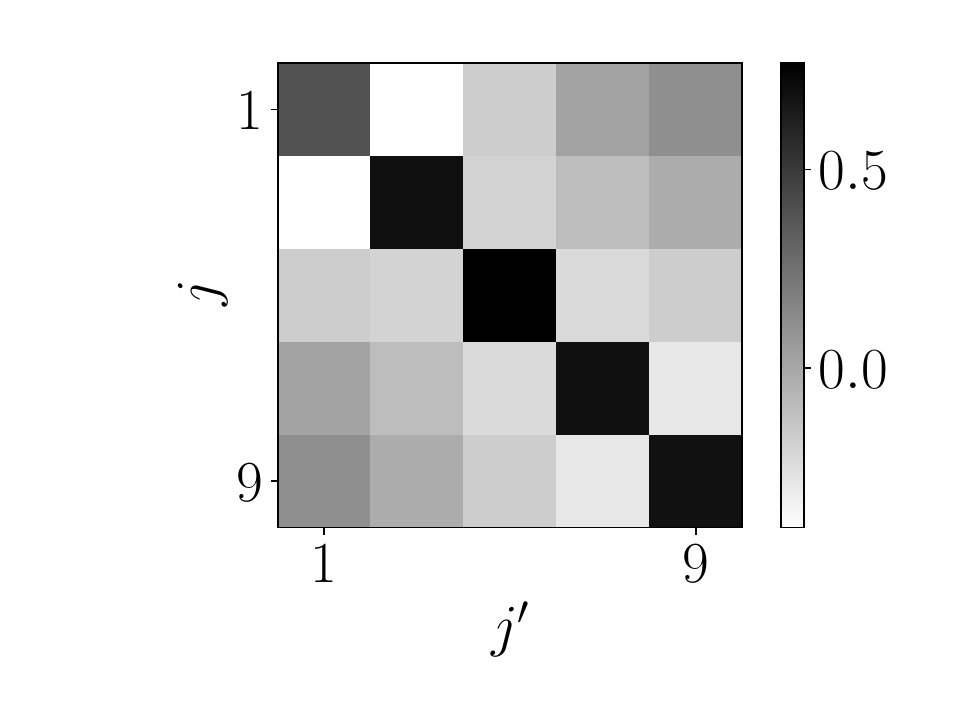}&
         \includegraphics[width=0.32\textwidth,trim={2.7cm 0.8cm 1.2cm 0.8cm},clip]{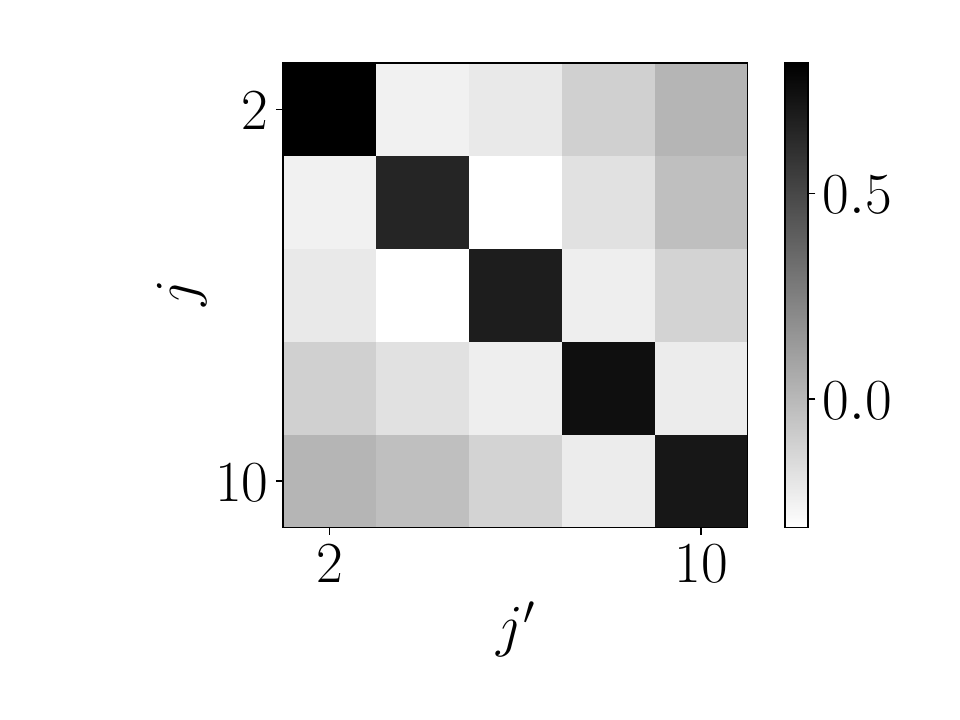}&
         \includegraphics[width=0.32\textwidth,trim={2.7cm 0.8cm 1.2cm 0.8cm},clip]{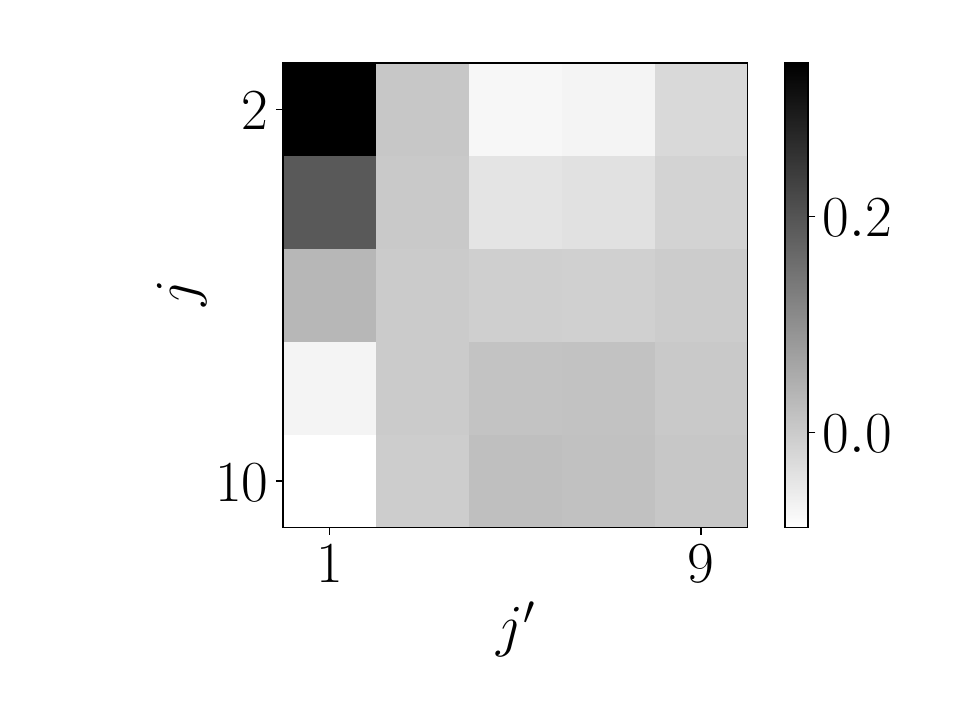}
\end{tabular}
        \caption{The values of the overlap matrix $\mathcal{Y}_{Pjm}^{Pj'm}(\theta_{\rm NA})$ defined by Eq.~(\ref{eq:CutedAngulInteg}) for $m=0$ and $P$ either $E$ or $M$
        are illustrated taking $j$ and $j^\prime$ as odd (even) values in the first (second) column. In the third column the overlap matrix $\mathcal{Y}_{Ejm}^{Mj'm}(\theta_{\rm NA})$ for different polarization is illustrated for $m=1$, such overlap is zero for $j$ and $j^\prime$ with the same parity. The upper (lower) row corresponds to a numerical aperture 0.75 (0.9).  }

\label{fig:ynnpP}

\end{figure*}

In the first (second) column of Fig.~\ref{fig:ynnpP}, the values of the overlap matrix $\mathcal{Y}_{Pj0}^{Pj'0}(\theta_{\rm NA})$
are illustrated for $j$ and $j^\prime$ even (odd) values and  numerical apertures 0.75 and 0.90; they are identical for  SVW modes with the same, either electric or magnetic, polarities.
In the last column of the same figure the values of the overlap matrix between electric ASVW and magnetic ASVW, $\mathcal{Y}_{Ejm}^{M^\prime j'm}(\theta_{\rm NA})$ is illustrated; $m=1$; $j$ and $j^\prime$ are chosen, and different parities are required to yield  values different from zero, Eq.~(\ref{eq:YEM}).

\begin{figure*}[hbt!]
     \centering
     \begin{subfigure}[b]{0.45\textwidth}
         \centering
         \includegraphics[width=\textwidth]{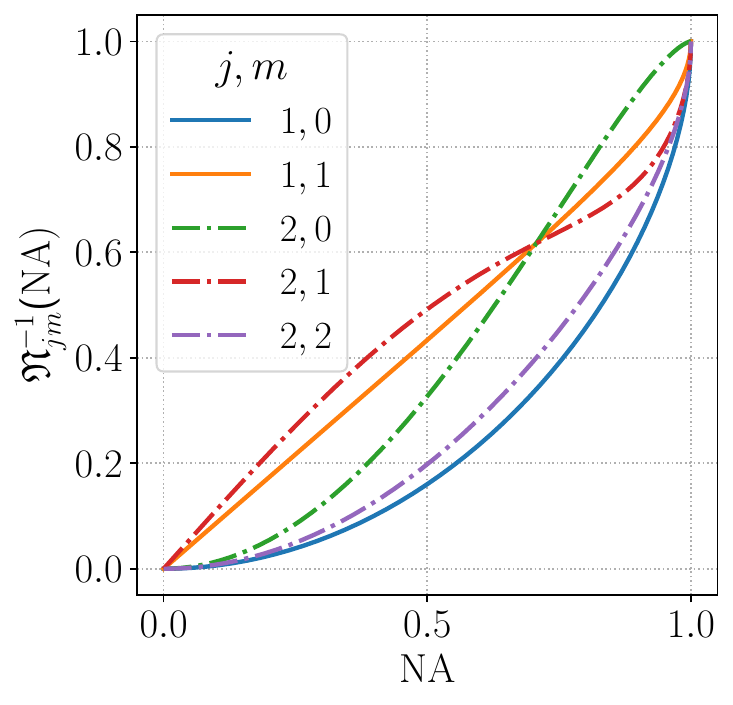}
         \caption{}
         \label{fig:VectorPotenNorma1}
     \end{subfigure}
     \hspace{0.01in}
     \begin{subfigure}[b]{0.45\textwidth}
         \centering
         \includegraphics[width=\textwidth]{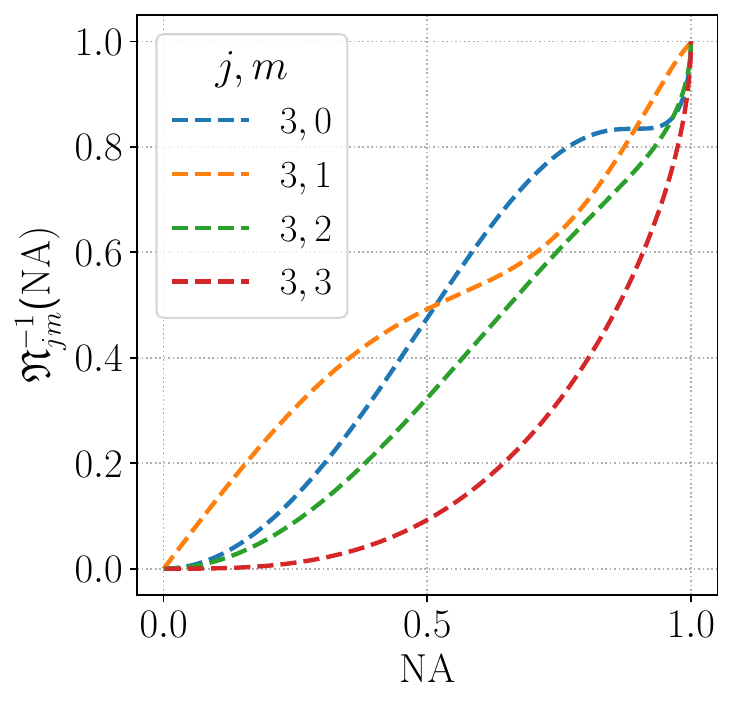}
         \caption{}
         \label{fig:VectorPotenNorma2}
     \end{subfigure}
        \caption{Normalization coefficients $\mathfrak{N}_{jm}^{-1}({\mathrm{NA}})$ for ASVW with $j\le 3$}
\label{fig:VectorPotenNorma}
\end{figure*}

Note that the diagonal elements of  $\mathcal{Y}_{Pjm}^{Pjm}(\theta_{\rm NA})$ could be used to define the angular spectrum of normalized approximated vectorial spherical
harmonics
\begin{equation}
\tilde{\mathbb{Y}}^{({\rm NA})}_{\nu}(\theta,\phi) = 
\begin{cases}\mathfrak{N}_{\nu}^{({\rm NA})} \tilde{\mathbb{Y}}^{(P)}_{jm}(\theta,\phi) \quad (\theta,\phi)\in \Omega_{{\rm NA}}\\ 0 \quad\quad\quad\quad\quad\quad\quad\quad (\theta,\phi)\notin \Omega_{{\rm NA}}
\end{cases}\label{eq:asASVW}
\end{equation}
that would satisfy the equation
\begin{equation}
\int_{\Omega_{{\rm NA}}}d\Omega \tilde{\mathbb{Y}}^{({\rm NA})*}_{\nu}(\theta,\phi)\cdot\tilde{\mathbb{Y}}^{({\rm NA})}_{\nu}(\theta,\phi) =1.
\end{equation}
The normalization coefficients are equivalent to the overlap of an ASVW with itself,
\begin{equation}
\vert \mathfrak{N}_{\nu}^{({\rm NA})}\vert^{-2} = \mathcal{Y}_{Pjm}^{Pjm}(\theta_{{
\rm NA}}).
\end{equation}
Figure \ref{fig:VectorPotenNorma} illustrates the behavior of the normalization term $[\mathfrak{N}_{jm}({\mathrm{NA}})]^{-1}$ as a function of the numerical aperture ${\mathrm{NA}}$ for $j\le 3$.

\
\subsection{Comparison between ASVW and SVW}

Under ideal conditions, the electromagnetic field of the approximate spherical vector modes would yield exact
spherical vector waves for a numerical aperture ${\rm NA}=1$. In this section we illustrate the behavior of the electric and magnetic fields for high {\rm NA}'s as they would result from Eq.~(\ref{eq:PlaneWaveDeco}). 

\begin{figure}[h]
     \centering
     \begin{subfigure}[b]{0.24\textwidth}
         \centering
         \includegraphics[width=\textwidth,trim={5cm 0 5.5cm 0},clip]{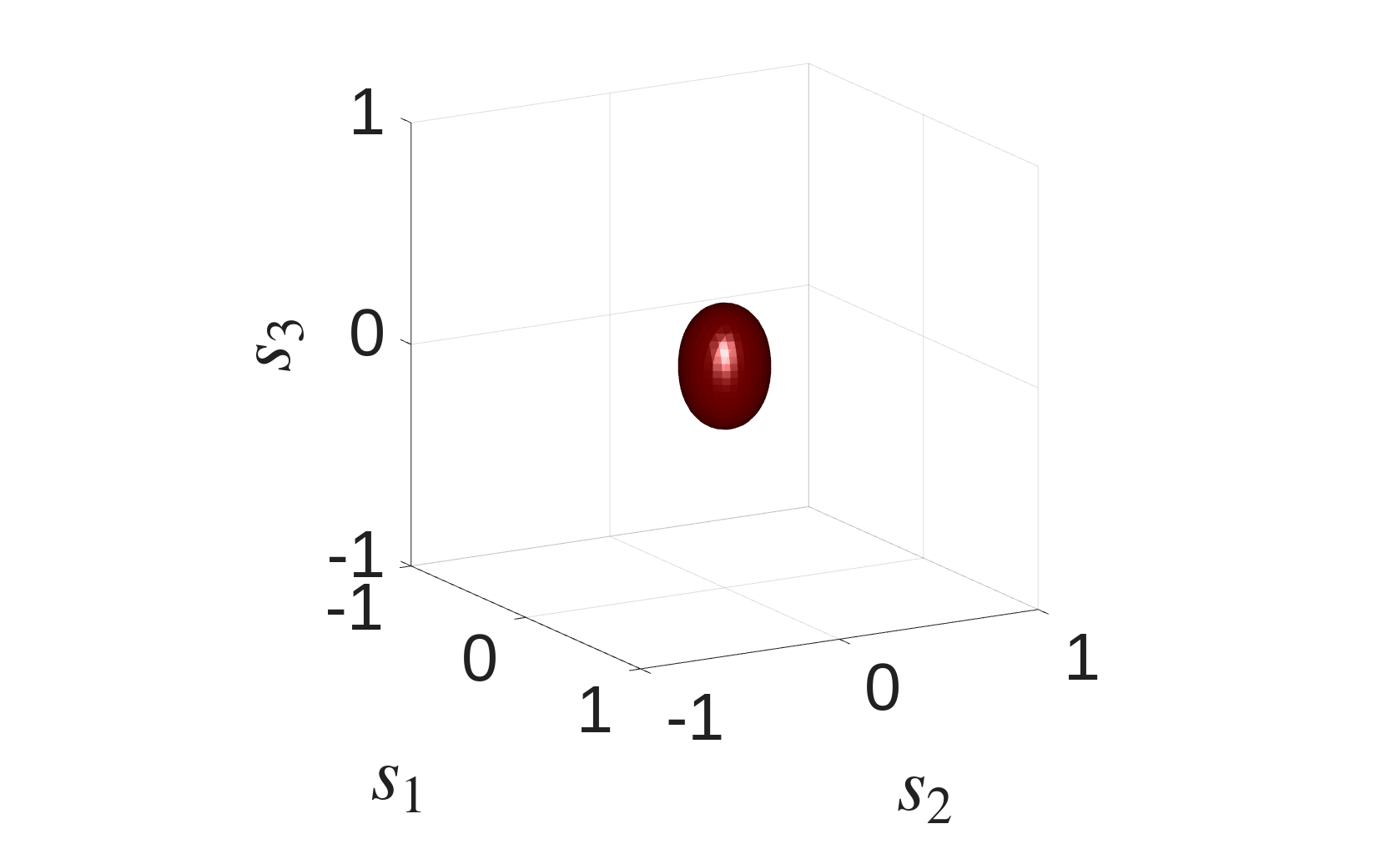}
         \caption{$\big|\mathbf{E}_{1,0}^{(E)}\big|^{2}$}
         \label{fig:YE10t_4piOA_NA1}
     \end{subfigure}
     \hspace{0.1cm}
     \centering
     \begin{subfigure}[b]{0.24\textwidth}
         \centering
         \includegraphics[width=\textwidth,trim={5cm 0 5.5cm 0},clip]{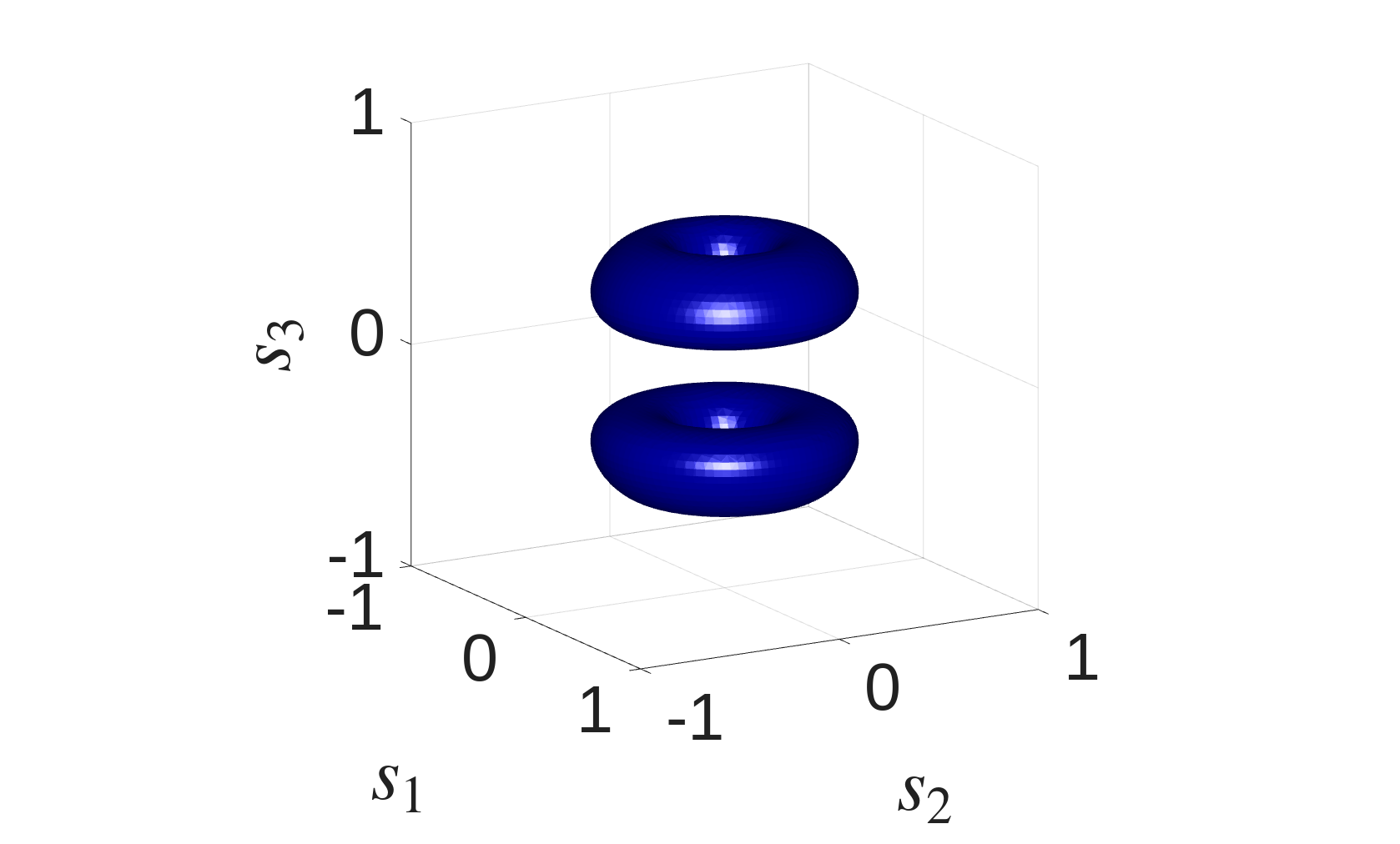}
         \caption{$\big|E_{1,0;\rho}^{(E)}\big|^{2}$}
         \label{fig:YE10r_4piOA_NA1}
     \end{subfigure}
    \hspace{0.1cm}
     \centering
     \begin{subfigure}[b]{0.24\textwidth}
         \centering
         \includegraphics[width=\textwidth,trim={5cm 0 5.5cm 0},clip]{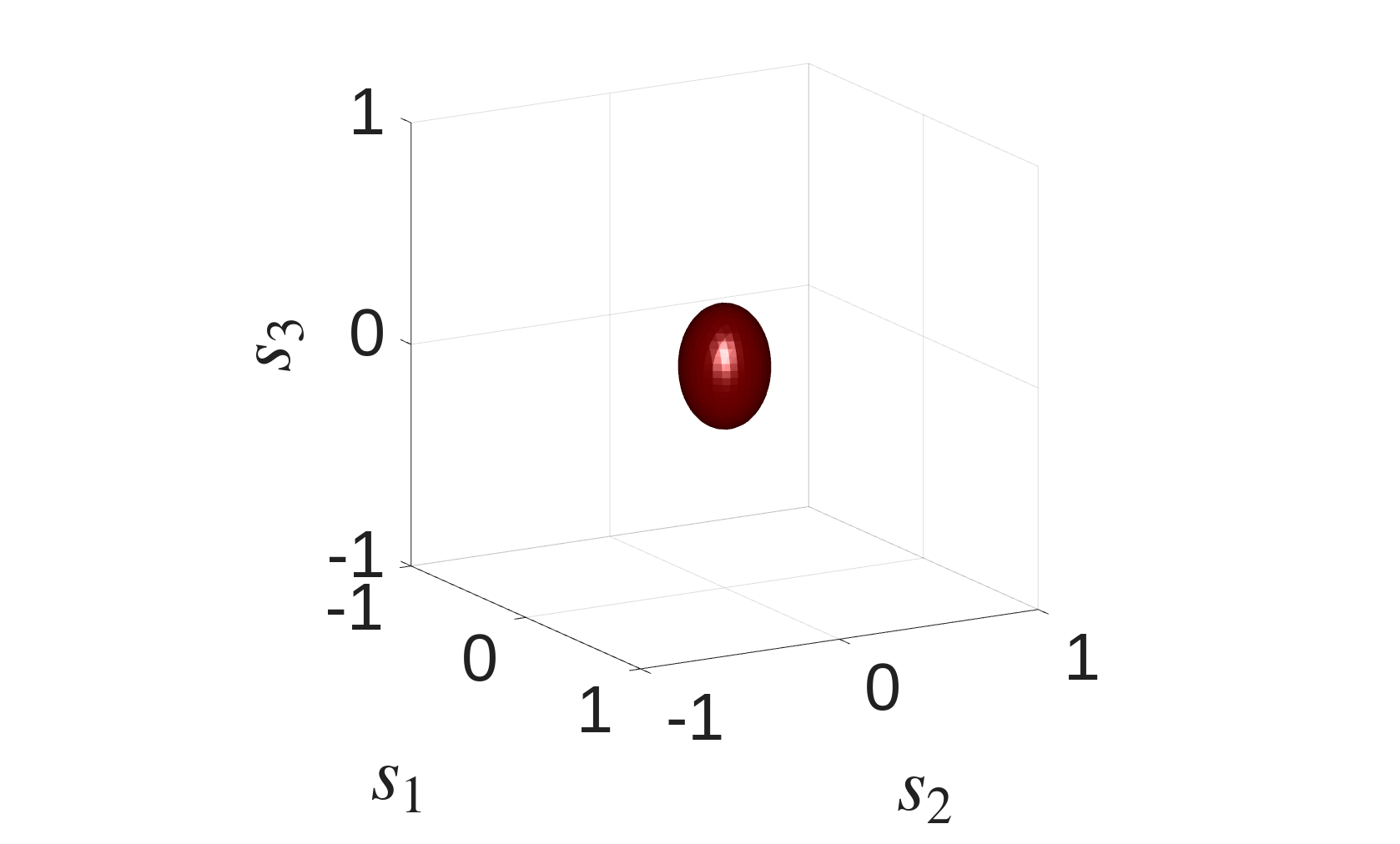}
         \caption{$\big|E_{1,0;z}^{(E)}\big|^{2}$}
         \label{fig:YE10z_4piOA_NA1}
     \end{subfigure}
\\
     \begin{subfigure}[b]{0.24\textwidth}
         \centering
         \includegraphics[width=\textwidth,trim={5cm 0 5.5cm 0},clip]{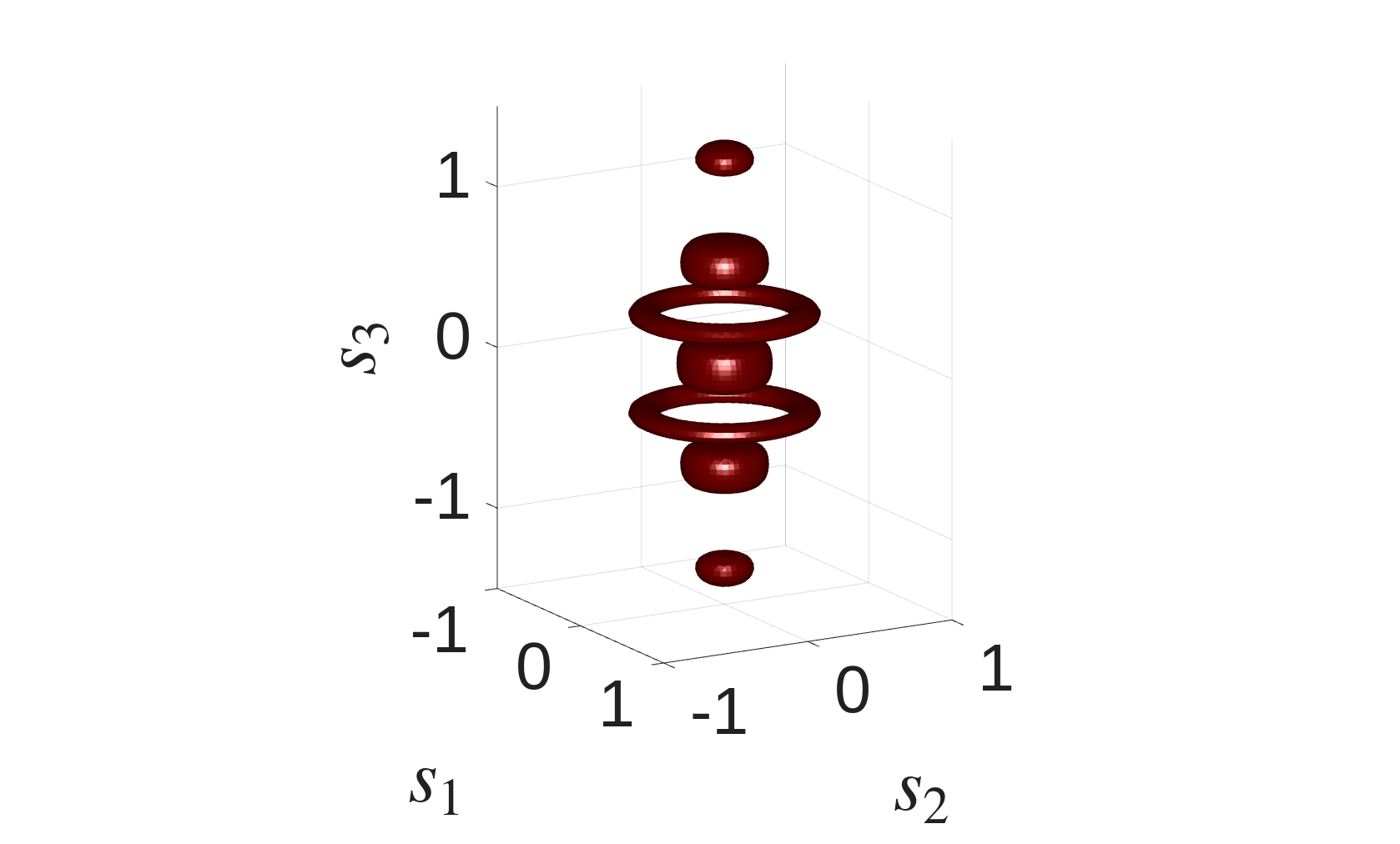}
         \caption{$\big|\mathbf{E}_{1,0}^{(E)}\big|^{2}$}
         \label{fig:YE10t_4piOA_NA0.75}
     \end{subfigure}
     \hspace{0.1cm}
     \centering
     \begin{subfigure}[b]{0.24\textwidth}
         \centering
         \includegraphics[width=\textwidth,trim={5cm 0 5.5cm 0},clip]{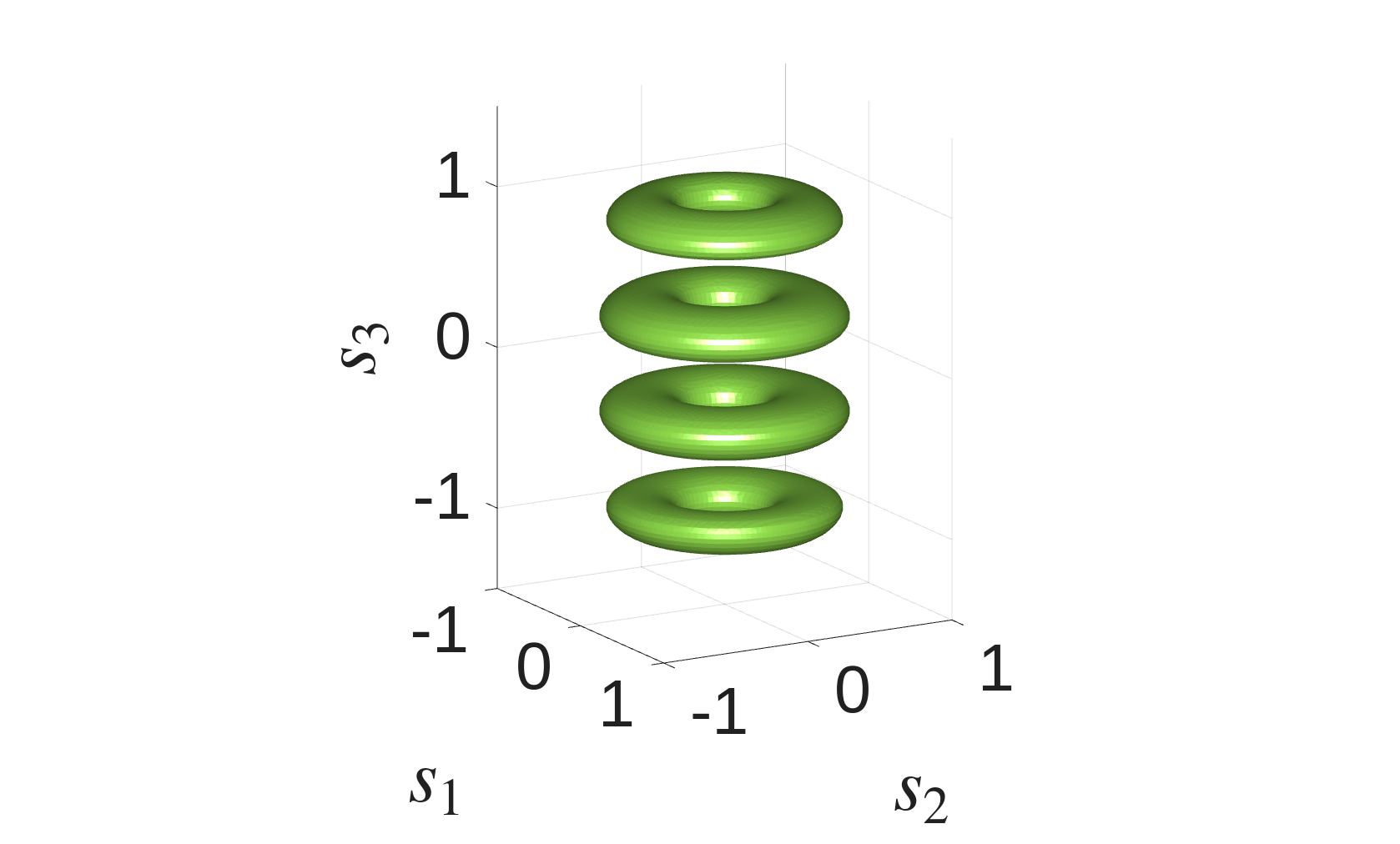}
         \caption{$\big|E_{1,0;\rho}^{(E)}\big|^{2}$}
         \label{fig:YE10r_4piOA_NA0.75}
     \end{subfigure}
    \hspace{0.1cm}
     \centering
     \begin{subfigure}[b]{0.24\textwidth}
         \centering
         \includegraphics[width=\textwidth,trim={5cm 0 5.5cm 0},clip]{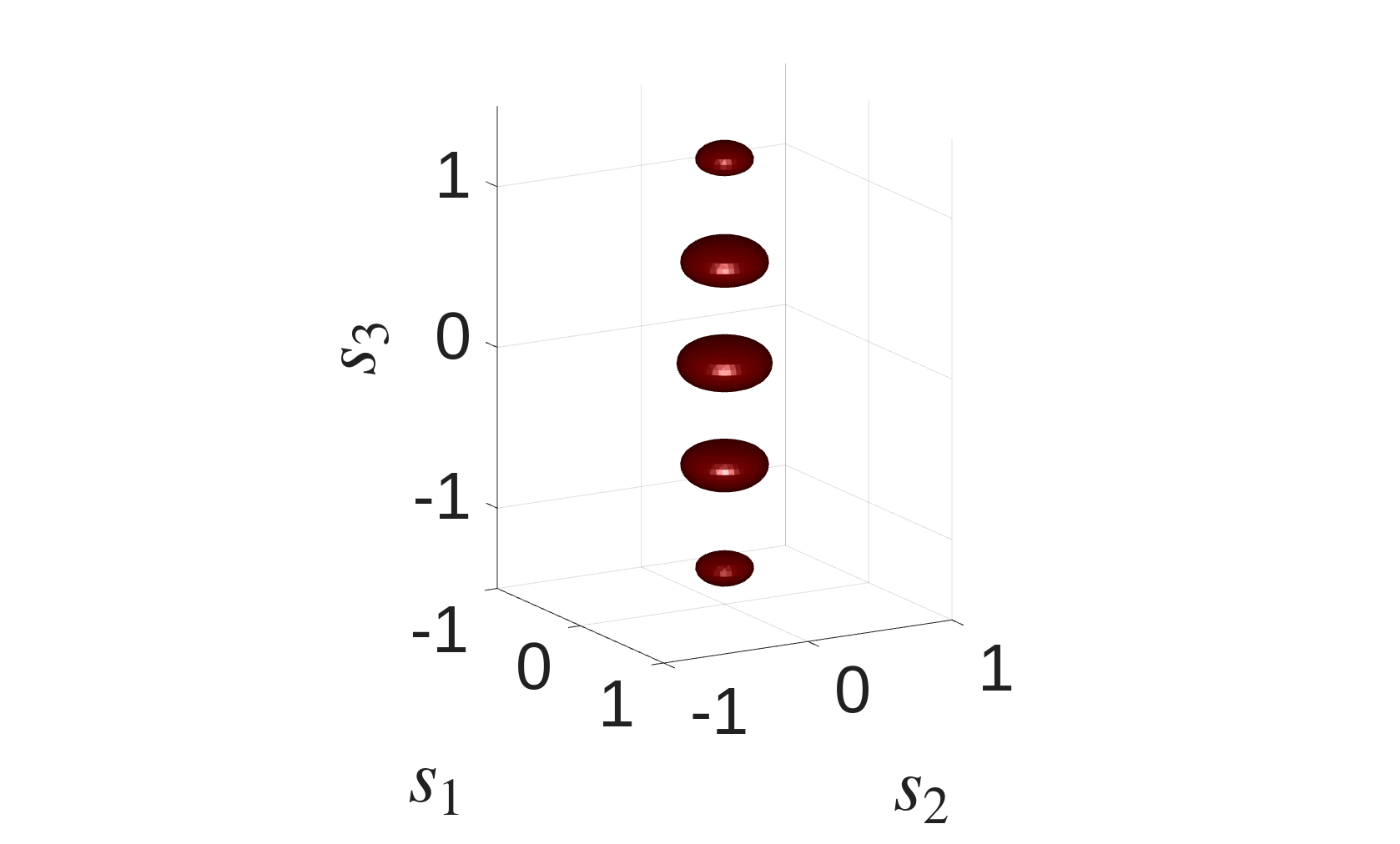}
         \caption{$\big|E_{1,0;z}^{(E)}\big|^{2}$}
         \label{fig:YE10z_4piOA_NA0.75}
     \end{subfigure}
\\
     \begin{subfigure}[b]{0.5\textwidth}
         \centering
         \includegraphics[width=\textwidth,trim={0 0 0 0},clip]{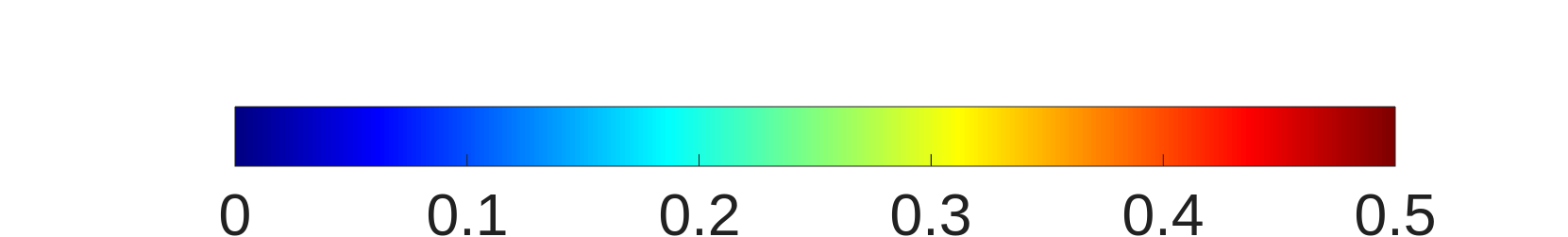}
     \end{subfigure}
        \caption{Isointensity surfaces ($0.5I_{\max}$) for the focused field of Eqs.~(\ref{eq:AnuNA},\ref{eq:asASVW}) associated to the electric field $\mathbf{E}_{1,0}^{(E)}(\mathbf{r})$ of an approximate  spherical vector wave;  (a, d) refer to the total field, (b, e) to the radial component and (c,f) to the longitudinal component of $\mathbf{E}_{1,0}^{(E)}(\mathbf{r})$ 
        for an ideal numerical aperture $\mathrm{NA}=1$ for (a-c) and $\mathrm{NA}=0.75$ for (d-e); $\mathbf{s} = \mathbf{r}/\lambda$. }
        \label{fig:YE10IsonInt_4piOA}
\end{figure}

\begin{figure}[h]
     \centering
     \begin{subfigure}[b]{0.24\textwidth}
         \centering
         \includegraphics[width=\textwidth,trim={5cm 0 5.5cm 0},clip]{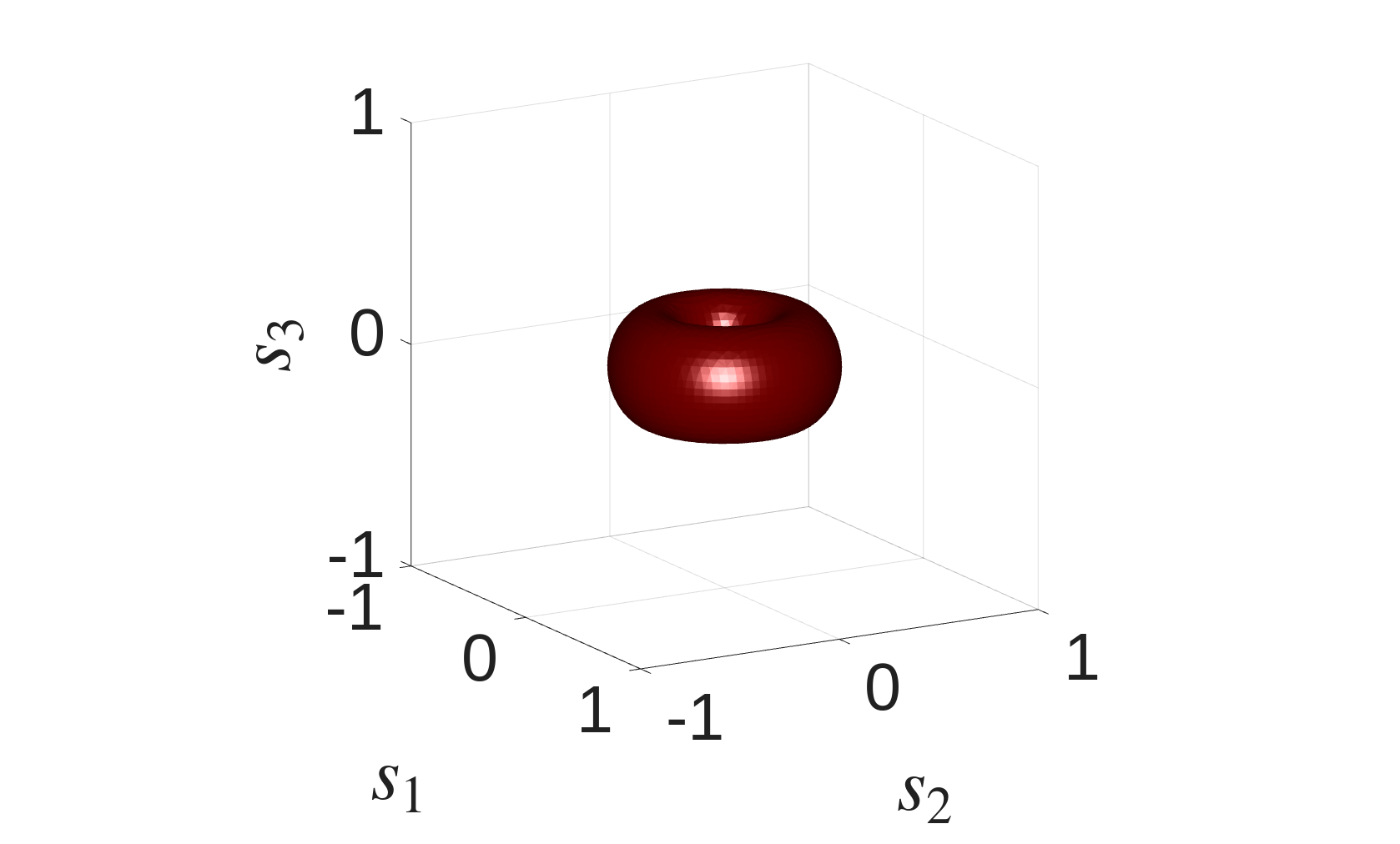}
         \caption{$\big|E_{1,0;\phi}^{(M)}\big|^{2}$}
         \label{fig:YM10p_4piOA_NA1}
     \end{subfigure}
     \hspace{0.1cm}
     \centering
     \begin{subfigure}[b]{0.24\textwidth}
         \centering
         \includegraphics[width=\textwidth,trim={5cm 0 5.5cm 0},clip]{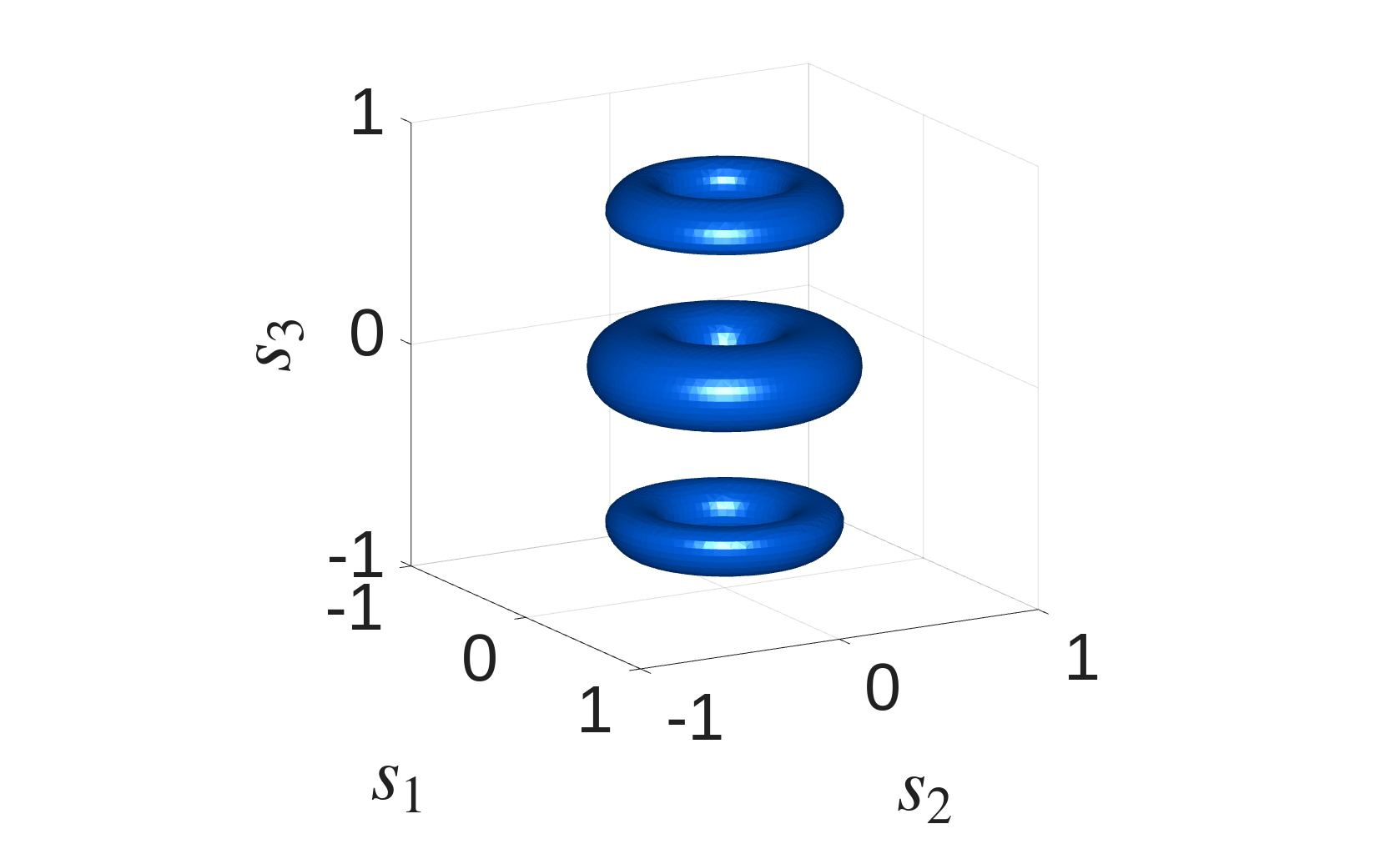}
         \caption{$\big|E_{1,0;\phi}^{(M)}\big|^{2}$}
         \label{fig:YM10p_4piO_NA0.9}
     \end{subfigure}
    \hspace{0.1cm}
     \centering
     \begin{subfigure}[b]{0.24\textwidth}
         \centering
         \includegraphics[width=\textwidth,trim={5cm 0 5.5cm 0},clip]{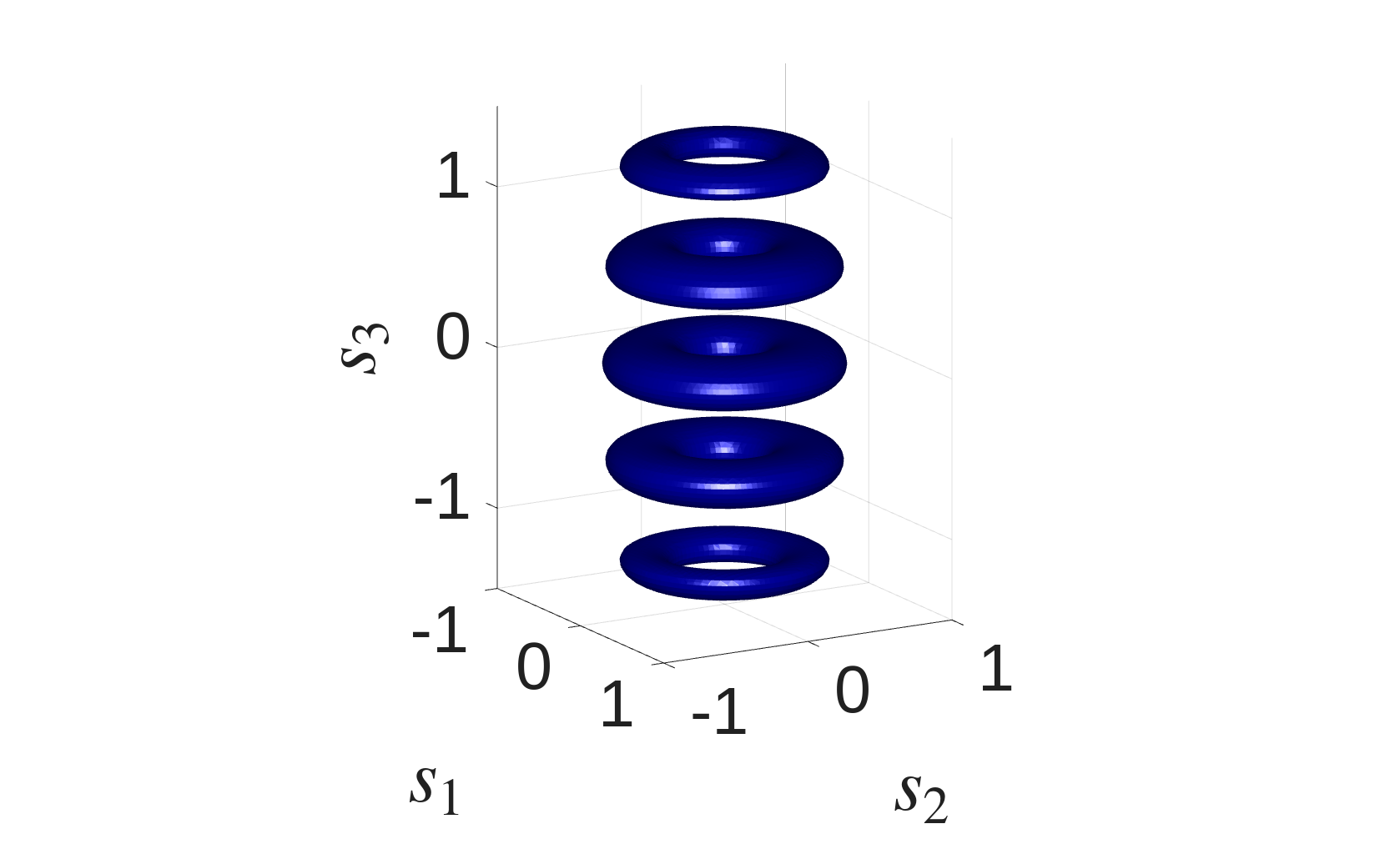}
         \caption{$\big|E_{1,0;\phi}^{(M)}\big|^{2}$}
         \label{fig:YM10p_4piOA_NA0.75}
     \end{subfigure}
\\
     \begin{subfigure}[b]{0.5\textwidth}
         \centering
         \includegraphics[width=\textwidth,trim={0 0 0 0},clip]{{JetCb}.png}
     \end{subfigure}
        \caption{Isointensity surfaces ($0.5I_{\max}$) for the azimuthal component of the magnetic 
        $\mathbf{E}_{1,0}^{(M)}(\mathbf{r})$,(i) $\mathrm{NA}=1$, (ii) $\mathrm{NA}=0.9$ and (iii) $\mathrm{NA}=0.75$;  $\mathbf{s} = \mathbf{r}/\lambda$.}
        \label{fig:YM10IsonInt_4piOA}
\end{figure}

This comparison is carried out by illustrating
the intensities of the electric fields at short distances to the focus of the array. Note that, in general, close to a radiation source
the multipole pattern of the EM field differs significantly from that at the radiation region, i.e., at large distances from the source. Here short distances are specially relevant because an atom close to the origin responds to the EM field at such a location. A quantitative comparison in terms of the fidelity between SVW and ASVW was already reported in Ref.~\cite{aquino2023}.
In Figure \ref{fig:YE10IsonInt_4piOA} the isointensity surfaces corresponding to half the highest intensity value are
shown for the  associated  electric field of approximate and ideal electric dipole SVW $\mathbf{E}_{1,0}^{(E)}(\mathbf{r})$.
In the first row  the ideal numerical aperture $\mathrm{NA}=1$ is considered while in the second row $\mathrm{NA}=0.75$.
The first column refers to  the total electric field, the second to the intensity of the radial component and the third column  to the intensity of the longitudinal component of $\mathbf{E}_{1,0}^{(E)}(\mathbf{r})$.  In Figure~\ref{fig:YM10IsonInt_4piOA}, the intensity pattern is illustrated for an ASVW with magnetic polarization and angular momentum $j=1$.
In general, the 4$\pi$ array preserves the vorticity of the fields which is encoded in the $e^{im\phi}$ dependence of
the modes. Meanwhile, the finite aperture of the lens leads to the presence of several lobes in the symmetry axis replacing a single lobe in the ideal ${
\rm NA}=1$ SVW. This behavior is consistent with the observations carried out by Hell et al who reported a similar fragmentation of an incident Gaussian beam in their seminal study of the optics of a 4$\pi$ array \cite{Hell94}. It can also be recognized that the number of fragmentation lobes increases as the numerical aperture decreases, they have a subwavelength structure and their separation ($\sim 0.5\lambda$ in the illustrated cases) also depends on the numerical aperture. Note the remarkable similarity of the central lobe structure to that of the SVW near the focus.

\subsection{Quantization of the EM field using monochromatic  ASVW modes }

The quantization of the EM field using ASVW modes must be based on the scalar product
$$\langle \sigma^\prime;{\rm NA}\vert \sigma;{\rm NA}\rangle =:\frac{1}{ 4\pi} \int d^3{\mathbf{r}}\big( \mathbf{E}_{\sigma^\prime;{\rm NA}}^*( \mathbf{r},t) \cdot \mathbf{E}_{\sigma;{\rm NA}}( \mathbf{r},t)
+  \mathbf{B}_{\sigma^\prime;{\rm NA}}^*( \mathbf{r},t) \cdot \mathbf{B}_{\sigma;{\rm NA}}( \mathbf{r},t)\big).
$$ 
In the wave-vector space  it reads
$$
\langle \sigma^\prime;{\rm NA}\vert \sigma;{\rm NA}\rangle = $$
\begin{equation}
\frac{\delta(k-k^\prime)}{ 4\pi k^2} \int_{\Omega_{{\rm NA}}} d\Omega\Big[
 \tilde{\mathbf{E}}_{\sigma^\prime;{\rm NA}}^*( \theta_{\hat{\mathbf{n}}}, \phi_{\hat{\mathbf{n}}}) \cdot \tilde{\mathbf{E}}_{\sigma;{\rm NA}}( \theta_{\hat{\mathbf{n}}}, \phi_{\hat{\mathbf{n}}})
+   \tilde{\mathbf{B}}_{\sigma^\prime;{\rm NA}}^*( \theta_{\hat{\mathbf{n}}}, \phi_{\hat{\mathbf{n}}}) \cdot \tilde{\mathbf{B}}_{\sigma;{\rm NA}}( \theta_{\hat{\mathbf{n}}}, \phi_{\hat{\mathbf{n}}})  \Big] .\label{eq:spASVW}
\end{equation}
The magnetic $\tilde{\mathbf{B}}_{\sigma;{\rm NA}}$ and electric $\tilde{\mathbf{E}}_{\sigma;{\rm NA}}$ angular spectra  are directly obtained from the  vector potential spectrum $\tilde{\mathbf{A}}_{\sigma;{\rm NA}}$. In general, those fields are built as a {\bf superposition} of the elementary ASVWs  with  angular spectrum Eq.~(\ref{eq:PlaneWaveDeco}). A direct calculation that takes into account Eq.~(\ref{eq:EBBE}), shows that for  elementary modes
\begin{equation}
\langle k^\prime P^\prime j^\prime m^\prime ;{\rm NA}\vert k P j m ;{\rm NA}\rangle = \frac{2\delta(k-k^\prime)}{ 4\pi c^2} \delta_{PP^\prime}\delta_{mm^\prime}\mathcal{Y}_{Pjm}^{Pj'm}(\theta_{{\rm NA}})
\end{equation}
The behavior of the overlap integral $\mathcal{Y}_{Pjm}^{Pj'm}(\theta_{{\rm NA}})$ allows the identification of sets of modes  $\{\vert \nu_0;{\rm NA}\rangle\}$ with orthogonal elements
 obtained by superposing elementary modes  which share their frequency $\omega$, polarization $P$, projection of total angular momentum along the symmetry axis $m$ and parity $\mathfrak{p}$. Such sets involve an additional label $n$ that result from a chosen orthogonalization procedure. That is modes of each set have labels
$$\bar{\sigma}\, \epsilon\,\{\omega,P,m,\mathfrak{p},n\}.$$
 Taking the parity as a good quantum number guarantees that the polarization label $P$ is also a good quantum number since
the matrix  $\mathcal{Y}_{\nu}^{\nu'}(\theta_{{\rm NA}})$ is zero if $P=E$, $P^\prime =M$ and their parity $\mathfrak{p}$ coincides.

The Gram-Schmidt procedure  singularizes a particular mode and consider superpositions where elementary modes are  added one by one with superposition coefficients that guarantee orthogonality with previous modes. The particular mode would be labeled with $n=1$ which would simultaneously correspond to a particular $j$ value.

 An alternative orthogonalization procedure corresponds to diagonalize  the overlap matrix $\mathcal{Y}_{Pjm}^{Pj'm}(\theta_{{\rm NA}})$ for a set of elementary ASVW that is considered to include the most relevant ones under the physical conditions of interest. Then, $n$ would be a  numbering label of the corresponding eigenvectors.

Finally, the obtained orthogonal modes would be normalized according to Einstein quantization algorithm
\begin{equation}
\langle k \sigma;{\rm NA}\vert k^\prime \sigma^\prime;{\rm NA}\rangle  = \hbar \omega \frac{\delta(k - k^\prime)}{k^2}\delta_{P,P^\prime}\delta_{mm^\prime}\delta_{\mathfrak{p},\mathfrak{p}^\prime}\delta_{nn^\prime}.
\end{equation}

\subsubsection{Illumination with an ASVW ($\theta_{{\rm NA}}\in(0,\pi/2$))}

The quantization of the EM field with ASVW is essential for the understanding of the spontaneous radiative decay of atomic systems or the quantum statistics of the light scattered by them. 
In this Section we consider an atomic gas that is located nearby the focus of the 4$\pi$ array and is illuminated using  incoming laser fields with a frequency, polarization and phase structure that makes reasonable to describe the EM field as a  coherent state of a particular elementary ASVW mode with an amplitude $\mathcal A_{\bar{\sigma};{\mathrm NA}}$,
\begin{equation}\label{eq:EMVecPotComPmode2}
\mathbf{A}_{\bar{\sigma};{\mathrm NA}}(\mathbf{r};\theta_{{\mathrm{NA}}}) = \mathcal A_{\bar{\sigma};{\mathrm{NA}}} \mathbb{Y}_{\omega\nu}(\mathbf{r};\theta_{{\mathrm{NA}}}),
\end{equation}
The formalism reported in Section 3 can then be applied to identify the immediate response of the atoms to this illumination scheme. 
The $\mathbf{A}_{\bar{\sigma}}(\mathbf{r};\theta_{{\mathrm{NA}}})$ mode is then written as a superposition of
SVW, Eqs.(\ref{eq:sup}-\ref{eq:oksn}). The coefficients $\mathcal{O}(k;\sigma,\nu)$ are determined by the overlap matrix $\mathcal{Y}_{\sigma}^{\nu'}(\theta_{{\mathrm{NA}}})$, Eq.~(\ref{eq:CutedAngulInteg}),
\begin{eqnarray}
\mathcal{O}(k;\sigma,\nu) 
&=&
\int d\Omega_{\mathbf{n}}\tilde{\mathbb{Y}}^{(P)*}_{jm}(\mathbf{n})\cdot\widetilde{\boldsymbol{\mathcal{A}}}_{\bar{\sigma};{\mathrm{NA}}}(\mathbf{n})
\\&=&
\mathcal A_{\bar{\sigma};{\mathrm{NA}}}
\mathcal{Y}_{\sigma}^{\nu}(\theta_{{\mathrm NA}}).
\end{eqnarray}
The resulting expression for the transition matrix probabilty for the stimulated emission or absorption of SVW photons depends directly on the multipole moments  $\big(Q_{\nu}(k_0)\big)_{fi}$ of the transition current $j_{fi;\mu}$  and on the illuminating mode $\boldsymbol{\mathcal{A}}_{\bar{\sigma};{\mathrm NA}}$ according to the equation
\begin{equation}
w_{fi}^{\bar{\sigma}_{0}} = \frac{4\pi^{2}}{\omega_{0}}  
\vert \mathcal A_{\bar{\sigma};{\mathrm{NA}}}\vert^2\Big|\sum_{\nu}\mathcal{Y}_{\sigma}^{\nu}(\theta_{{\mathrm{NA}}}) \big(Q_{\nu}(k_0)\big)_{fi}\Big|^{2}.\label{eq:wASVW}
\end{equation}
As a consequence, the transition probabilities can be manipulated via the numerical aperture and the amplitude $\mathcal A_{\bar{\sigma};{\mathrm{NA}}}$ of the ASVW. The latter is determined by the amplitude of the incoming EM field in Regions I and III of the 4$\pi$ array. Note that Eq.~(\ref{eq:wASVW}) makes also evident the possibility of interference effects on the transition probability $w_{fi}^{\bar{\sigma}_{0}}$. The feasibility of observing such effects is determined by the relative value of the spherical multipole moments $Q_{\nu}(k_0)$ of the transition current ${\mathbf{j}}_{fi}$.

\begin{figure*}[htb!]
     \centering
     \begin{subfigure}[b]{0.27\textwidth}
         \centering
         \includegraphics[width=\textwidth]{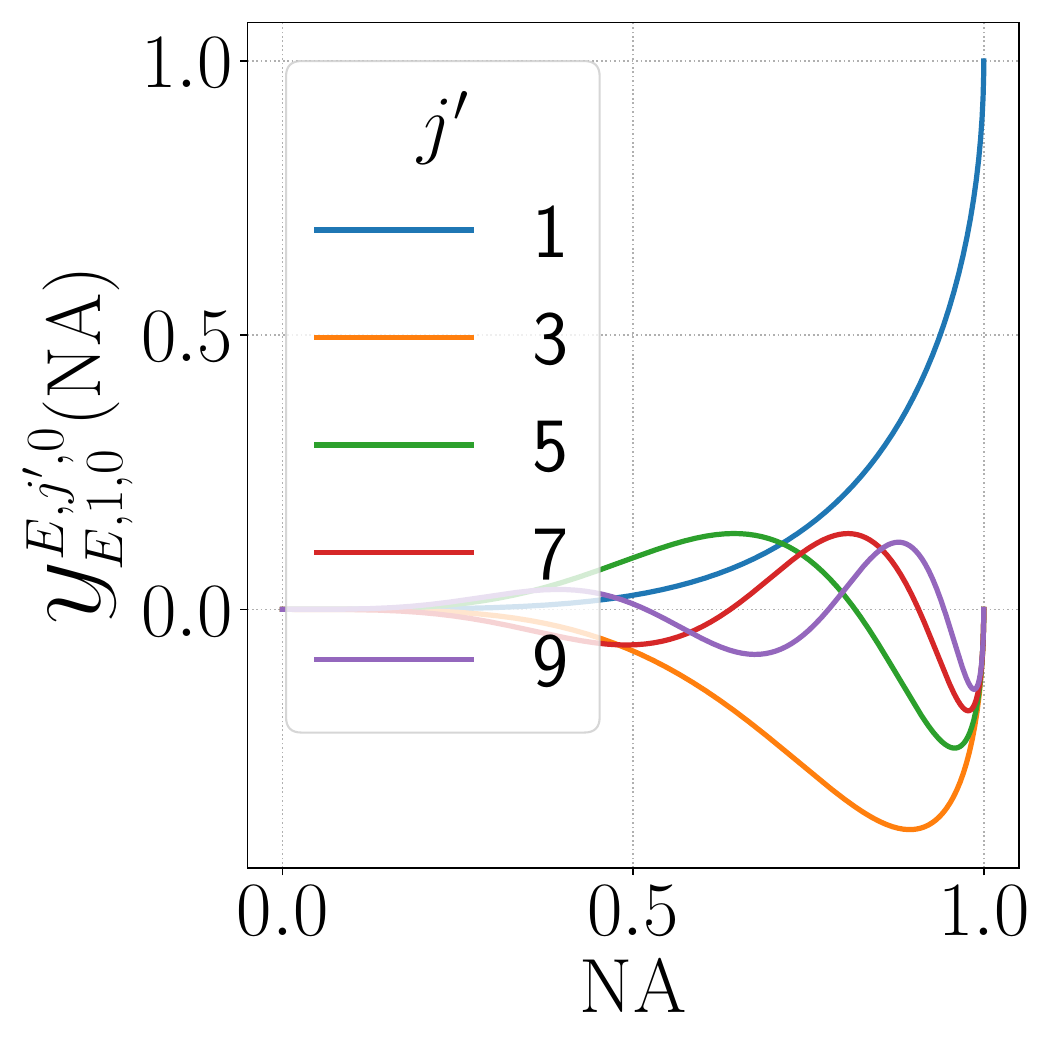}
         \caption{}
     \end{subfigure}
     \hspace{0.01in}
     \begin{subfigure}[b]{0.27\textwidth}
         \centering
         \includegraphics[width=\textwidth]{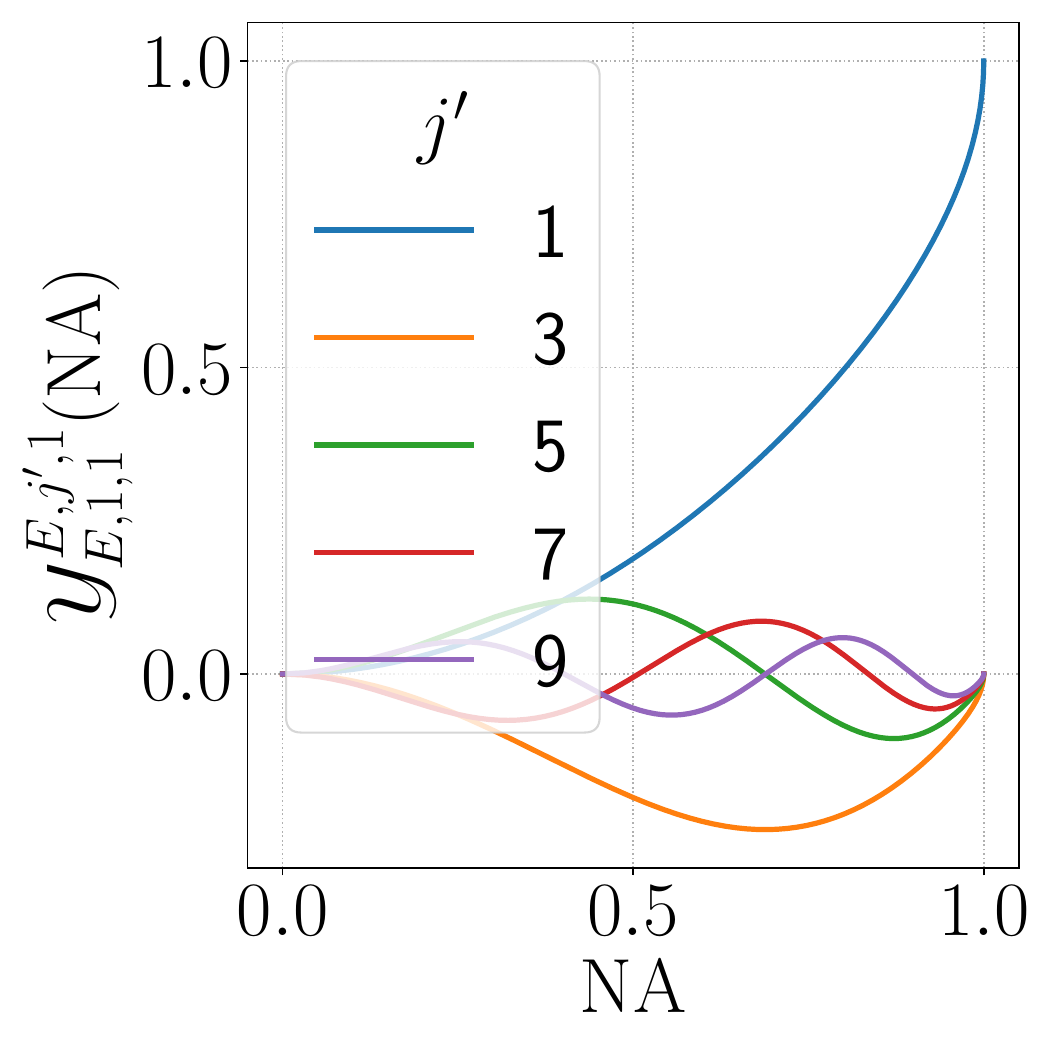}
         \caption{}
     \end{subfigure}
     \hspace{0.01in}
     \begin{subfigure}[b]{0.27\textwidth}
         \centering
         \includegraphics[width=\textwidth]{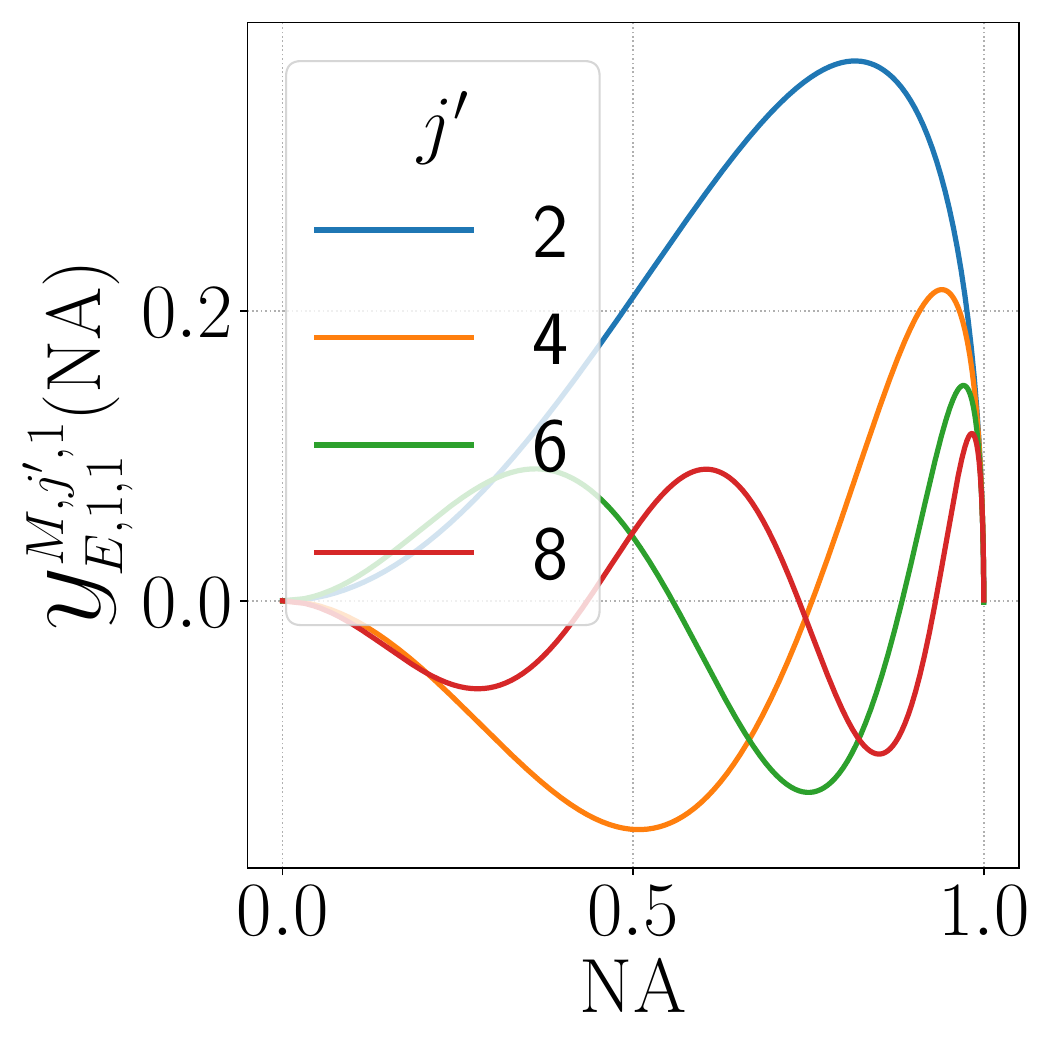}
         \caption{}
     \end{subfigure}
        \caption{Overlap factor  $\mathcal{Y}_{\sigma}^{\nu'}(\theta_{\mathrm {NA}})$ as a function of the numerical aperture for an electric ASVW for $j=1$ and (a)  $m=0$ and (b) $m=1$, with an electric ASVW  with $j^\prime =1,3,5,7,9$. Subfigure (c) illustrates the overlap factor between an electric ASVW with $j=1$ and $m=1$  and
        a magnetic ASVW with $j^\prime=2,4,6,8$.}
\label{fig:TraPro_Hatom_E-ASVW}
\end{figure*}

In Figure~\ref{fig:TraPro_Hatom_E-ASVW}, the behavior of the overlap factor  
$\mathcal{Y}_{\sigma}^{\nu}(\theta_{{\mathrm{NA}}})$ as a function of the numerical aperture is illustrated for an electric ASVW with $j=1$ with  $m=0$ and with $m=1$. It can be observed that, in general, 
$\mathcal{Y}_{\sigma}^{\nu}(\theta_{{\mathrm{NA}}})$ is not a monotonic function of {\rm{NA}}. Besides the overlap with SVWs with magnetic polarization is not null and, for a quadrupole magnetic wave, it exhibits a significant maximum value at a feasible {\rm NA}. This result could be useful in the study of magnetic quadrupole transitions using optimized properties of incoming elementary ASVW. Note that an accurate calculation of magnetic multipole transition rate requires the incorporation of spin effects~\cite{Trautmann23}. However, the modular character of our formalism accounts for the structure of the illuminating field via the simple  effective coupling encoded in the overlap matrix $\mathcal{Y}_{\sigma}^{\nu}(\theta_{{\rm NA}})$ element.

\section{Conclusions and Perspectives}

The one-to-one correspondence of spherical multipole moments and the SVW emitted or absorbed by an atomic system can be used  to design mechanisms to control the transition rates of atomic systems. The formalism introduced in this work is a useful tool to that end. It just requires knowledge of the overlap matrix between the involved EM modes with SVW which plays a similar role to that of the density of states function. 
 
In the specific case of approximate spherical vector waves generated in a 4$\pi$ optical array, we have shown that the numerical aperture of the lens is a control parameter for the amplification or inhibition of stimulated emission or absorption photons. This holds true even for incident photons with a polarization (electrical or magnetic) different from that of the corresponding spherical vector wave. 

The efficiency of stimulated transitions is expected to be greater than that based on illumination with high-intensity paraxial beams. For example, electrical quadrupole transitions are generally described by an effective coupling of the Cartesian electric quadrupole moment $Q_{rs}$ to the gradient of the local electric field $\mathbf{E}$, given by $Q_{rs}\partial_r{\mathbf{E}}_s$ \cite{mojica2017}. For a paraxial beam, the electric field gradient has its maximum value along the principal direction of propagation, with a value $\sim k E$ where $k$ is the field wavenumber and $E$ its amplitude. The ASVW gradient near the focus is maximum in the radial direction and a natural scale that is also $\sim k E$. One advantage provided by ASVW is the additional presence of optical vortices for $m\ne 0$ that resemble those of quadrupole electrical SVW.

The impressive advances in controlling the internal and external degrees of atoms using electromagnetic fields in recent years also increase the feasibility of observing the phenomena predicted in this work. Individual atoms can be cooled and trapped at the focus of the 4$\pi$ array with nanometric uncertainties. Current lens objectives achieve numerical apertures greater than 0.9.

So far, we have focused on atomic transition rates. Modifying the electromagnetic energy landscape of the center of mass motion of the atom opens a different avenue. If the frequency of an illuminating ASVW is properly selected, the spatial dependence of the light field induces an effective potential for the center of mass motion of the atoms. Standard optical tweezers result from the electric dipole EM fields coupling to the electric dipole moment for far of resonance illumination. The structured light landcape of ASVW
generated by a 4$\pi$ optical array could be used to generate optical traps with interesting structures based on multipole moments. The intensities, detuning and  numerical apertures would be control
parameters for the dynamics of the center of mass of the atoms.

In summary, the effective coupling of an atomic system to the electromagnetic field via electric dipole transitions is perhaps one of the most fruitful approaches. However, recent advances in the generation of structured light fields and the feasibility of trapping individual atoms in well-localized spatial regions open the possibility of increasing the role of alternative multipole moments of matter in the development of studies in both basic science and quantum engineering. ASVW is a promising tool for that purpose.

\end{document}